\documentclass[prl,twocolumn,aps,showpacs,amssymb,superscriptaddress]{revtex4-1}

\usepackage{epsfig}
\usepackage{dcolumn}
\usepackage{amssymb}
\usepackage{color}

\begin{document}

\title{\bf Systematic investigation of the effects of disorder at the lowest order \\ throughout the BCS-BEC crossover}

\author{F. Palestini}
\affiliation{LENS, Universit\`{a} di Firenze, 50019 Sesto Fiorentino (FI), Italy}
\affiliation{Division of Physics, School of Science and Technology, Universit\`{a} di Camerino, 62032 Camerino (MC), Italy}

\author{G. C. Strinati}
\affiliation{Division of Physics, School of Science and Technology, Universit\`{a} di Camerino, 62032 Camerino (MC), Italy}
\affiliation{INFN, Sezione di Perugia, 06123 Perugia (PG), Italy}

\pacs{03.75.Ss,05.30.Jp,72.15.Rn,05.60.Gg}

\begin{abstract}
A systematic investigation of the effects of disorder on the BCS-BEC crossover at the lowest order in the impurity potential is presented for the normal phase above the critical temperature $T_{c}$.
Starting with the t-matrix approach for the clean system, by which pairing correlations between opposite-spin fermions evolve from the weak-coupling (BCS) to the strong-coupling (BEC) limits by increasing the strength of the attractive inter-particle interaction, \emph{all\/} possible diagrammatic processes are considered where the effects of a disordered potential are retained in the self-energy at the lowest order.
An accurate numerical investigation is carried out for all these diagrammatic terms, to determine which of them are mostly important throughout the BCS-BEC crossover.
Explicit calculations for the values of $T_{c}$, the chemical potential, and the Tan's contact are carried out.
In addition, the effect of disorder on the single-particle spectral function is analyzed, and a correlation is found between an increase of $T_{c}$ and a widening of the pseudo-gap energy at $T_{c}$ on the BCS side of unitarity in the presence of disorder, while on the BEC side of unitarity the presence of disorder favors the collapse of the underlying Fermi surface.
The present investigation is meant to orient future studies when the effects of disorder will be considered at higher orders, 
with the purpose of limiting the proliferation of diagrammatic terms in which interaction and disorder are considered simultaneously.
\end{abstract}

\maketitle
\section{I. Introduction} 
\label{sec:introduction}

The interplay between interaction and disorder has long been of interest in condensed-matter physics, especially in the context of the metal-insulator 
transition.
Its theoretical treatments have proved highly non-trivial and require the use of rather sophisticated field-theoretical and diagrammatic methods 
\cite{LR-1985,BK-1994}.
Pioneering work in this sense was done in Refs.\cite{Finkelstein-1982,CD_CML-1984}.
Simplifying assumptions used in that context rested on the presence of a ``large'' Fermi surface (related to the underlying Fermi-liquid description of metals \cite{AGD-1963}) and on addressing mostly ``universal'' properties, which are related to critical phenomena and for which the system details that are poorly under control are not fully relevant.

More recently, ultra-cold Fermi gases have opened the possibility of performing an accurate experimental control of the system parameters, specifically of the inter-particle interaction (that can be varied almost at will via the use of Fano-Feshbach resonances) and of the external potential in which the atoms are embedded (via suitable arrangements of laser fields) \cite{Varenna-2007}.
In general, Fermi as well as Bose ultra-cold gases are then regarded as ``quantum simulators'', which allow one to realize a variety of models that would otherwise defy an accurate control with more conventional condensed-matter systems.

For these reasons, interest has raised in exploiting ultra-cold gases to address problems that have remained unaccomplished with more conventional disordered materials, like the occurrence of the Anderson localization due to disorder when the inter-particle interaction is artificially switched off 
both in Bose \cite{Aspect-2008,Inguscio-2008} and in Fermi \cite{De_Marco-2011} systems, or even the emergence of the coherent backscattering for weaker disorder \cite{Aspect-2012}.
At present, experimental interest is raising to study the interplay between disorder and interaction in the context of the BCS-BEC crossover, whereby the pairing correlations due to the attractive inter-particle interaction can be varied in a continuous fashion from weak to strong coupling.
Questions like the localization of a Cooper pair as a whole, or its disrupt by disorder that would localize the two fermions independently, are of much interest also for disordered superconductors \cite{Ioffe-2011}.
In addition, an advantage of an attractive interaction over a repulsive one is that it cannot lead by itself to localization effects in the absence of disorder. 

Despite the considerable experimental efforts devoted to the problem, there appear to be limited theoretical achievements thus far in exploring the interplay of disorder and interaction throughout the BCS-BEC crossover.
These include the work of Ref.\cite{Orso-2007}, where the effects of weak disorder on the BCS-BEC crossover were considered at low temperature
in the superfluid phase using a functional integral with a Gaussian action for the bosonic fluctuations over and above the BCS mean field. 
As well as the work of Ref.\cite{Sa_de_Melo-2011}, that addressed the value of critical temperature $T_{c}$ for the superfluid transition from the normal phase in the presence of weak disorder across the BCS-BEC crossover, also using a functional integral formulation.
In both these references, diagrammatic processes were not explicitly identified at the fermionic level, a practice that would instead help one to describe the physical processes where disorder and interaction act at the same time, as well as the way they evolve from the weak (BCS) to the strong (BEC) coupling limits.

Purpose of the present work is to study in a \emph{systematic} way how a system of fermions with strong pairing correlations, which evolves throughout the BCS-BEC crossover, is affected by the presence of disorder when this is treated at the lowest significant order.
This will be done by identifying \emph{all} possible diagrammatic processes through which the disorder affects directly the fermionic single-particle self-energy, whereby the inter-particle interaction is treated at the level of the t-matrix that has been extensively used to describe the BCS-BEC crossover for a clean system in the normal phase \cite{Randeria-1993,PPPS-2004}. 
All these self-energy diagrams will be calculated numerically from weak to strong coupling, and their contributions to the critical temperature and chemical potential will be explicitly analyzed.
In this way, a detailed control will be achieved on those diagrammatic terms that are not only dominant in either the weak or strong coupling limits (a feature that will be checked independently by analytic estimates as well), but also contribute in an appreciable way in the unitary limit of most interest which is intermediate between the BCS and BEC limits.
By a further comparative analysis of the single-particle spectral function with and without disorder, an \emph{increase} of $T_{c}$, that we shall find on the BCS side of unitarity in the presence of disorder, will be related to a corresponding increase of the pseudo-gap energy at $T_{c}$.

The main issues addressed in this paper and the key new physical results can be briefly summarized as follows:

\noindent
(i) The first main issue is about the \emph{role of the underlying Fermi surface} when the scattering by impurities is combined with an attractive pairing interaction spanning the 
BCS-BEC crossover. To this end, standard approximations, which rely on the presence of a ``large'' Fermi surface and are commonly used in the theory of disordered electronic systems, will have to be abandoned.

\noindent
(ii) The second main issue stems from the fact that, since the BCS-BEC crossover has thus far been explicitly realized with ultra-cold Fermi gases, \emph{the disorder potential should be suitably modeled} to be as close as possible to the experimental realizations that can be done for these systems. Our choice of the impurity potential will thus be made in order to adapt this need to the use of a diagrammatic approach for dealing with the combined effects of interaction and disorder.

\noindent
(iii) The third main issue concerns \emph{the identification of a minimal set of relevant physical processes} at the level of the fermionic self-energy, where disorder at the lowest order is combined with interaction effects at the level of the t-matrix. This will be done through both analytic estimates that are separately possible in the BCS and BEC limits, and accurate numerical calculations that cover the whole BCS-BEC crossover. These processes will, in fact, be the only ones that should be dressed at higher orders in the disorder in future work, so as to avoid an unmanageable proliferation of terms when the disorder will be considered at infinite order (in terms, e.g., of disorder ladders and/or cross ladders) to reach eventually the regime where localization effects take place.

\noindent
(iv) One of the main physical results obtained by the present analysis is that the existence of an underlying Fermi surface on the BCS side of unitarity largely protects the system from the effects of disorder through Pauli blocking which limits the amount of impurity scattering processes. In this context, we shall find that \emph{disorder can actually favor the occurrence of pairing correlations}, with a simultaneous increase of $T_{c}$ and of the Tan's contact and a widening of the pseudo-gap energy.
This finding, that holds for an attractive inter-particle interaction, can be regarded to parallel to some extent that of Ref.\cite{Finkelstein-1994} for a repulsive inter-particle interaction, whereby the presence of disorder leads to an enhancement of the effects of the Coulomb repulsion among electrons.

\noindent
(v) A second important and related physical result is that, when the Fermi surface eventually collapses on the BEC side of unitarity and thus it can no longer protect the system from the effects of disorder, the system becomes essentially \emph{bosonic in nature and thus much more sensitive to the presence of disorder.} This result indicates that the evolution and fate of the underlying Fermi surface, which is the driving element behind the BCS-BEC crossover also in the clean case, acquires an even more marked relevance in the presence of disorder since this amplifies its effects on the system coherence.

\noindent
(vi) An additional main physical result is that, in the BEC limit when composite bosons form out of fermion pairs, a nice \emph{mapping can be established} between the diagrammatic structures for composite bosons which we recover from our analysis and for point-like bosons whose internal structure is immaterial.

For completeness, it is also worth mentioning that a connection between pairing fluctuations and disordered effects was also considered at the diagrammatic level in 
Ref.\cite{Chen-Schrieffer-2002}, albeit with the use of a different pairing theory (that was built on a quasi-two-dimensional single-band Hamiltonian in a lattice to make contacts with the physics of the cuprates) but with essentially no reference to the physics of the BCS-BEC crossover.
For these reasons, no one of the issues and physical results (i)-(vi) listed above were addressed or discussed in Ref.\cite{Chen-Schrieffer-2002}.

The paper is organized as follows. 
Section II describes the diagrammatic approach that we adopt for the BCS-BEC crossover in the presence of weak disorder, and discusses the assumptions underlying the treatment of disorder plus interaction in conventional condensed-matter systems which are, however, going to break down when departing from the weak-coupling limit. 
Analytic results are also reported in the weak- as well as in the strong-coupling limits.
For the latter, a mapping is further provided to the self-energy of non-interacting composite bosons in the presence of disorder.
In Section III the numerical calculations for the critical temperature $T_{c}$ of the normal-superfluid transition and for the corresponding chemical potential $\mu$ throughout the BCS-BEC crossover are presented, together with the results for the Tan's contact which is of special interest to the physics of ultra-cold atoms since it englobes a number of universal properties of systems with short-range dynamics \cite{Tan-2008,Braaten-2012}. 
Section IV addresses the characterization of the single-particle spectral function for the relevant fermionic and bosonic excitations in their respective regimes, and discusses a correlation found between the increase of $T_{c}$ and of the pseudo-gap energy on the BCS side of unitarity.
Section V gives our conclusions and outlines future lines of research dealing with stronger disorder.
The Appendix reports details on the average over the disorder that we have adopted. 

\vspace{-0.2cm}
\section{II. Diagrammatic approach} 
\label{sec:diagrammatic_theory}

\begin{figure}[t]
\begin{center}
\includegraphics[angle=0,width=9.0cm]{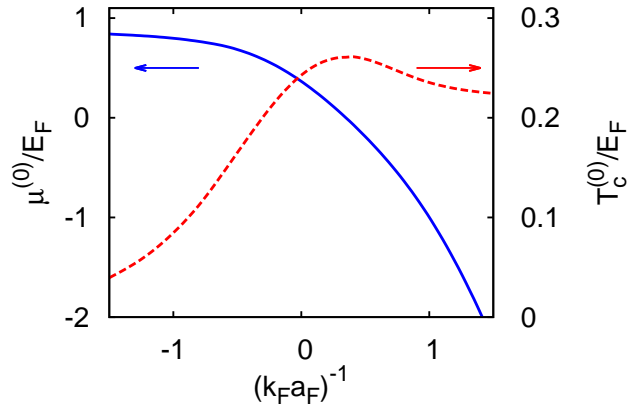}
\caption{Chemical potential $\mu^{(0)}$ (full line, left scale) calculated at the critical temperature $T_{c}^{(0)}$ (dashed line, right scale) 
              vs the coupling parameter $(k_{F} a_{F})^{-1}$. 
              Both $\mu^{(0)}$ and $T_{c}^{(0)}$ are in units of $E_{F}$ and are obtained in the absence of disorder using the t-matrix approximation 
              of Figs.÷\ref{fig-2}(b) and \ref{fig-2}(c) (see below).}
\label{fig-1}
\end{center}
\end{figure}

A key role in the BCS-BEC crossover is played by the variation of the chemical potential $\mu$ from one limit to the other, since at zero temperature $\mu$ evolves from the value of the Fermi energy $E_{F} = k_{F}^{2}/(2m)$ in the extreme BCS limit of non-interacting fermions, to (minus one half of) the value $(m a_{F}^{2})^{-1}$ of the binding energy for the two-fermion problem in vacuum in the extreme BEC limit of non-interacting composite bosons.
In the process, the underlying Fermi surface is progressively washed out.
A similar behavior occurs also at finite temperatures $T \ll T_{F}$ where $T_{F}$ is the Fermi temperature, as shown in Fig.÷\ref{fig-1} for temperatures that remain a fraction of $T_{F}$.
Here, $m$ is the fermion mass, $k_{F} = (3 \pi^{2} n)^{1/3}$ the Fermi wave vector where $n$ is the total density (two equally populated fermion species with $\uparrow$ and $\downarrow$ spins are considered throughout), and $a_{F}$ is the scattering length for the two-fermion problem in vacuum.
Correspondingly, the dimensionless coupling parameter $(k_{F} a_{F})^{-1}$ varies from being $\ll -1$ in the BCS limit to being $\gg +1$ in the BEC limit, it vanishes at unitarity, and its magnitude is $\lesssim 1$ in the window where the crossover takes place.

Inclusion of disorder across the BCS-BEC crossover has thus to take unavoidably into account the progressive disappearance of what was the underlying Fermi surface in the BCS limit and played a major role in that limit.
To highlight how this effect evolves in practice, we begin by considering the simplest process through which the impurity potential acts on a Fermi system.

\vspace{0.1cm}
\begin{center}
{\bf A. The weak-coupling limit and the role of the Fermi surface}
\end{center}
\vspace{0.1cm}

\begin{figure}[t]
\begin{center}
\includegraphics[angle=0,width=6.5cm]{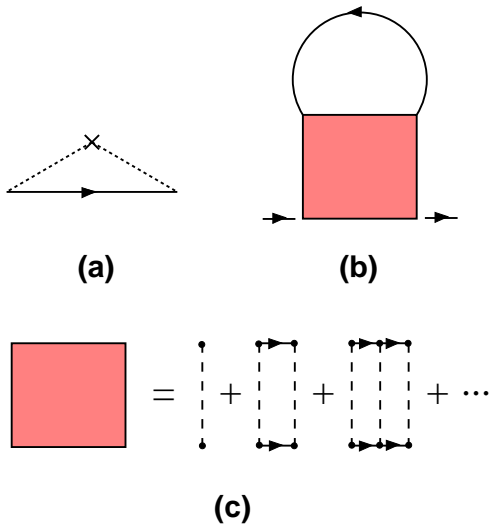}
\caption{Fermionic self-energy diagrams describing: (a) Non-interacting fermions affected by disorder at the lowest order; 
             (b) Interacting fermions at the level of the t-matrix in the absence of disorder; (c) t-matrix pair (ladder) propagator $\Gamma_{0}$ for the clean system. 
             Full, dashed, and dotted lines with a cross represent (bare) fermion propagators, inter-particle potential, and impurity scattering, respectively.}
\label{fig-2}
\end{center}
\end{figure}

For a system of non-interacting fermions, the self-energy contribution that takes into account the effect of disorder at the lowest order is depicted in
Fig.÷\ref{fig-2}(a), where the dotted line with a cross represents the scattering by the impurities \cite{AGD-1963-second}.

The averaging over the impurity configurations is performed as in Appendix A, with a procedure that is mostly suited to the present diagrammatic approach and where a Gaussian-correlated (white noise) disorder is recovered only as a limiting case. 
The analytic expression associated with the diagram of Fig.÷\ref{fig-2}(a) then reads ($\hbar = 1$ throughout):
\begin{equation}
\Sigma_{2a}(\mathbf{k},\omega_{n}) = \int \! \frac{d\mathbf{p}}{(2\pi)^{3}} \, \frac{u(\mathbf{p})^{2}}
{i \omega_{n} - (\mathbf{k+p})^{2}/(2m) + \mu} 
\label{simplest-self-energy}
\end{equation}
\noindent
where $\mathbf{k}$ and $\mathbf{p}$ are wave vectors, $\omega_{n}=(2n+1) \pi k_{B} T$ ($n$ integer) is a fermionic Matsubara frequency ($k_{B}$ being the Boltzmann constant).
Here, 
\begin{equation}
u(\mathbf{p})^{2} = \left\{   \begin{array}{ll}  \gamma &  \mbox{if $|\mathbf{p}| < p_{0}$}   \\
                                                                          0       &  \mbox{otherwise}
                                           \end{array}  \right.                            
\label{impurity-potential}
\end{equation}
\noindent
represents the (averaged) impurity potential where $\gamma$ is a constant and $p_{0}$ a wave-vector cutoff (with $p_{0} \gg k_{F}$ as discussed in Appendix A).

The reason to keep a finite (albeit large with respect to $k_{F}$) value of $p_{0}$ is evident when calculating the expression 
(\ref{simplest-self-energy}) in the BCS limit where $\mu \simeq E_{F}$ is the largest energy scale in the problem, such that $m \gamma p_{0} \ll E_{F}$ 
\cite{footnote-condition}.
One obtains for $|\mathbf{k}| \lesssim k_{F}$ and $|\omega_{n}| \ll E_{F}$:
\begin{equation}
\Sigma_{2a}(\mathbf{k},\omega_{n}) \simeq - \, \frac{m \, \gamma \, p_{0}}{\pi^{2}} 
                                                                               - i \, \gamma \, \frac{(2m)^{3/2}}{4 \pi} \,\sqrt{\mu} \,\, {\rm sgn}(\omega_{n}) 
\label{approximate-self-energy-BCS}
\end{equation}
\noindent
where the real part represents an energy shift at the lowest order in the impurity potential (which would diverge in the limit $p_{0} \rightarrow \infty$ of a truly Gaussian correlated potential).
In the theory of disordered metals, this term is usually dismissed as being an irrelevant constant that can be reabsorbed in a renormalization of the chemical potential \cite{AGD-1963-second}.
This is no longer possible in the context of the BCS-BEC crossover, for which the renormalization of the chemical potential is an essential ingredient of the problem and has thus to be explicitly considered.

The imaginary part of Eq.(\ref{approximate-self-energy-BCS}) can be expressed as $- \pi \gamma N_{0} \, {\rm sgn}(\omega_{n})$ in terms of the density of states $N_{0} = m k_{F}/(2 \pi^{2})$ per spin component at the Fermi level.
This result could have been obtained directly from the imaginary part of the expression (\ref{simplest-self-energy}) for which the integral is convergent even when  
$p_{0} \rightarrow \infty$.
In this case one can use the standard approximation \cite{AGD-1963}:
\begin{equation}
\int \! \frac{d\mathbf{k}}{(2\pi)^{3}} \, F\left(\frac{\mathbf{k}^{2}}{2m} - \mu\right) \simeq
N_{0} \, \int_{- \infty}^{+ \infty} \! d\xi \, F(\xi)
\label{approximate-integral}
\end{equation}
\noindent
that holds for a smooth function $F$  of $\xi(\mathbf{k}) = \mathbf{k}^{2}/(2m) - \mu$ for which the integral on the right-hand side is convergent.
The approximate way (\ref{approximate-integral}) to calculate the integrals over the wave vector has systematically been used in the theory of disordered electronic systems to simplify the calculations, whereby the presence of an underlying Fermi surface has invariably been assumed \cite{LR-1985,BK-1994}.

\begin{figure}[t]
\begin{center}
\includegraphics[angle=0,width=8.0cm]{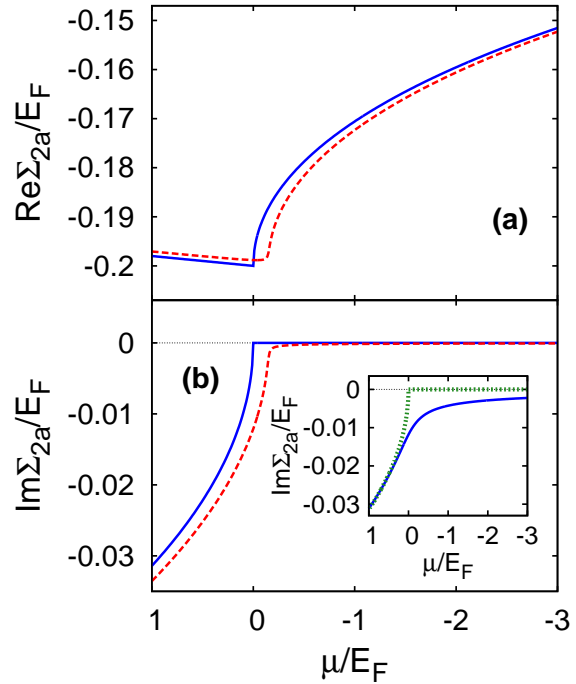}
\caption{Impurity self-energy $\Sigma_{2a}$ of Fig.÷\ref{fig-2}(a) calculated numerically for $\mathbf{k}=0$ vs the chemical potential $\mu$ over an extended   
              range of the BCS-BEC crossover, with disorder parameter $\tilde{\gamma} = 0.01$. 
             (a) Real and (b) imaginary part of $\Sigma_{2a}$ at $T = 0$ with $\omega_{n}=0^{+}$, where full and dashed lines correspond to the 
             non-self-consistent and self-consistent calculations.
             In addition, the inset of panel (b) shows the imaginary part of $\Sigma_{2a}$ with $\omega_{n}=\pi k_{B} T$ at $T = 0.1 T_{F}$, where full and dotted lines 
             correspond to the cases when the approximation (\ref{approximate-integral}) is avoided or adopted in the non-self-consistent calculation.}
\label{fig-3}
\end{center}
\end{figure}

This is \emph{no longer true} for the BCS-BEC crossover already past unitarity on the BEC side, and the approximation (\ref{approximate-integral}) has consequently to be abandoned.
In particular, in the (extreme) BEC limit where $\mu/E_{F} \rightarrow - \infty$ the value 
\begin{equation}
\Sigma_{2a}(\mathbf{k},\omega_{n}) \simeq \frac{\gamma \, p_{0}^{3}}{6\pi^{2}} \, \frac{1}{i \omega_{n} - \xi(\mathbf{k})} 
                                                  \simeq - \, \frac{\gamma \, p_{0}^{3}}{6\pi^{2} |\mu|} 
\label{approximate-self-energy-BEC}
\end{equation}
\noindent
of the expression (\ref{simplest-self-energy}) becomes irrelevant.

For a generic value of $\mu$ intermediate between the BCS and BEC regimes, the expression (\ref{simplest-self-energy}) can be calculated numerically and the result is reported in Figs.÷\ref{fig-3}(a) and \ref{fig-3}(b) for $\mathbf{k}=0$ and $\omega_{n}=0^{+}$.
[Here and in all following figures, the value of the disorder parameter $\gamma$ is given in terms of the dimensionless quantity 
$\tilde{\gamma}= \gamma \, m^{2}/(\pi^{2} k_{F})$.]
Also shown in the same figures is a more refined calculation which treats the fermion propagator of Fig.÷\ref{fig-2}(a) in a self-consistent way, and which results in a smoothing of the cusp near $\mu = 0$.
One sees from these plots that a marked change of behavior occurs at $\mu=0$ where the imaginary parts vanishes abruptly, thus resulting in the absence of any significant scattering by the impurities.
Additional diagrammatic contributions beside that of Fig.÷\ref{fig-2}(a) need thus to be included to study the effects of disorder on the BCS-BEC crossover, which we will consider in the following at the lowest relevant order.

It is interesting to note from Fig.÷\ref{fig-3}(b) that the expression (\ref{approximate-self-energy-BCS}) for the self-energy (with a characteristic square-root behavior of the imaginary part, as expected when the approximation (\ref{approximate-integral}) holds) remains valid provided $\mu$ is positive.
This is because in Fig.÷ \ref{fig-3}(b) we have reported the results for $\omega_{n}=0^{+}$ at zero temperature.
The difference between the results obtained by calculating the (imaginary part of the) self-energy $\Sigma_{2a}(\mathbf{k},\omega_{n})$ without and with the use of the approximation (\ref{approximate-integral}) becomes apparent for increasing $T$, as shown in the inset of Fig.÷\ref{fig-3}(b) where $\omega_{n}=\pi k_{B} T$ with 
$T = 0.1 T_{F}$ in the non-self-consistent calculation.
This is an additional indication that the approximation (\ref{approximate-integral}) cannot be used as soon as departing from the BCS limit of the crossover.

\vspace{0.1cm}
\begin{center}
{\bf B. Fermionic pairing self-energy terms at the lowest order in the disorder}
\end{center}
\vspace{0.1cm}

\begin{figure}[h]
\includegraphics[angle=0,width=8.5cm]{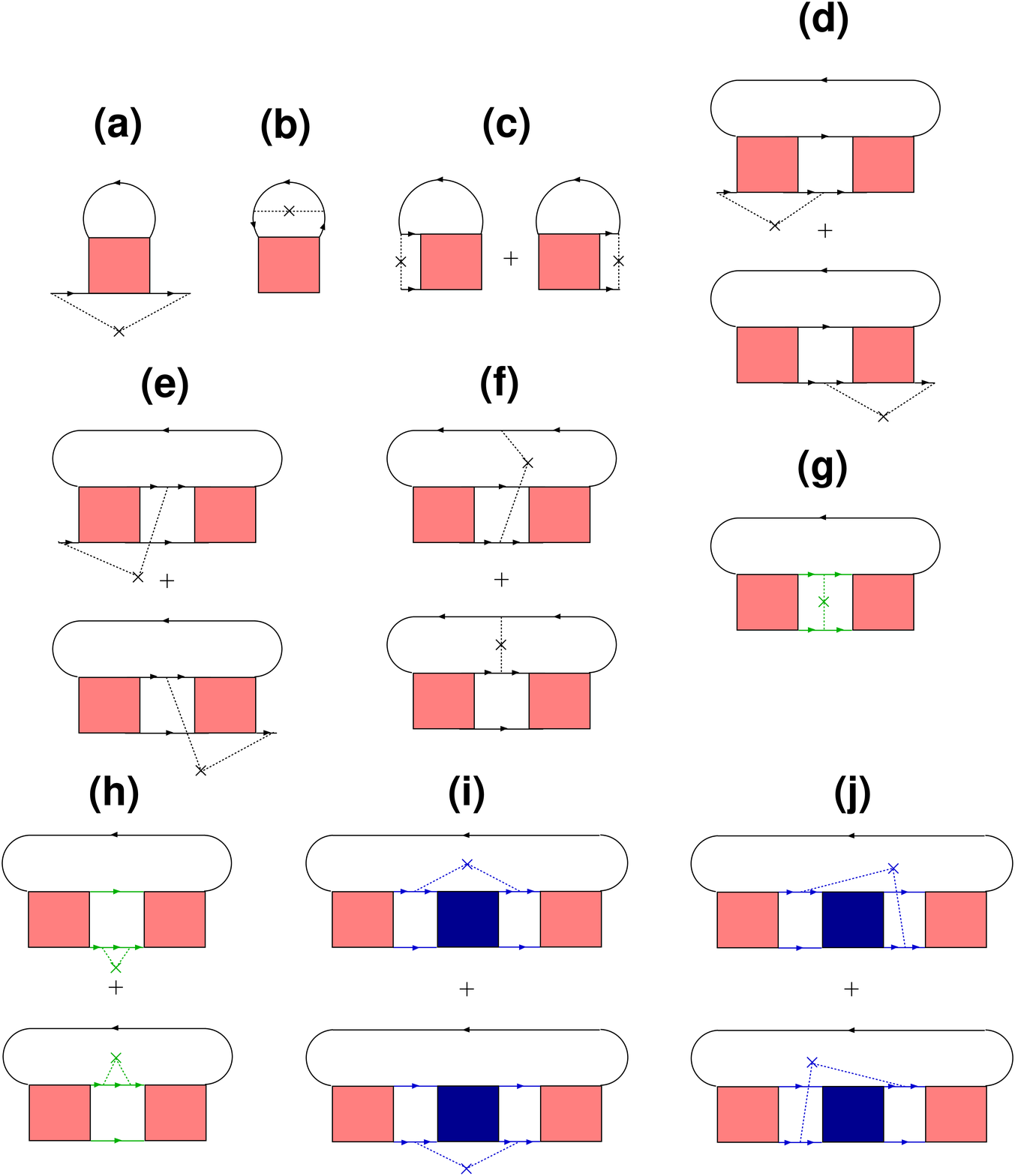}
\caption{Fermionic self-energy diagrams in the presence of disorder built on the t-matrix of Figs.÷\ref{fig-2}(b) and \ref{fig-2}(c), where all possible decorations 
              at the lowest order in the disorder have been inserted. Boxes stand for the ladder propagator $\Gamma_{0}$ of the clean system, full lines for the 
              bare fermion propagator $G_{0}$, and dotted lines with a cross for the impurity scattering.}
\label{fig-4}
\end{figure}

For a clean system, the BCS-BEC crossover can be described at finite temperature in the normal phase in terms of the fermionic self-energy of
Figs.÷\ref{fig-2}(b) and \ref{fig-2}(c) \cite{Strinati-12}.
The presence of disorder decorates this diagram in several ways, which are \emph{all} reported in Fig.÷\ref{fig-4} at the lowest order in the 
disorder \cite{footnote-additional_diagrams}.

A number of considerations can be made on these diagrams before they are explicitly calculated:

\noindent
(i) Diagrams \ref{fig-4}(a) and \ref{fig-4}(b) represent crossed self-energy insertions to the diagrams of Figs.÷\ref{fig-2}(a) and \ref{fig-2}(b), in the order.

\noindent
(ii) Diagrams \ref{fig-4}(c) are reminiscent of the corrections to the Hartree self-energy in the theory of disordered electronic systems \cite{LR-1985,BK-1994}.

\noindent
(iii) Diagrams \ref{fig-4}(d) and \ref{fig-4}(e) (which albeit topologically distinct have the same value when a contact inter-particle interaction is considered) represent processes where disorder affects the fermionic character of the system.
The same can be said of diagrams \ref{fig-4}(f). 

\noindent
(iv) Diagrams \ref{fig-4}(g) and \ref{fig-4}(h) contain the effect of disorder through a kind of ``bosonic'' self-energy insertion to the ladder propagator 
$\Gamma_{0}$ (as evidenced in the central part of the diagrams), which is constructed, however, in terms of fermionic quantities only.

\noindent
(v) Diagrams \ref{fig-4}(i) and \ref{fig-4}(j) (which, again, albeit topologically distinct have the same value for a contact inter-particle interaction) contain the effect of disorder through a truly ``bosonic'' self-energy insertion to the ladder propagator (as evidenced in the central part of the diagrams), that physically corresponds to a self-energy insertion for composite bosons (made up of a fermion pair) in the presence of disorder.

From the above analysis, one expects diagrams \ref{fig-4}(c) to be of importance in the BCS limit, and diagrams \ref{fig-4}(i) and \ref{fig-4}(j) to be of importance in the BEC limit.
This expectation will be verified below through analytic estimates of the diagrams in these limits, and will be checked by accurate numerical calculations in Section III.
Numerical calculations will further be required to determine the way the diagrams of Fig.÷\ref{fig-4} contribute to physical quantities (like the critical temperature and chemical potential), especially in the unitary region of most interest which is intermediate between the BCS and BEC limits.

It further turns out that, besides in diagram \ref{fig-4}(b), it is strictly necessary to keep a finite value of the cutoff $p_{0}$ also in diagrams \ref{fig-4}(g)-\ref{fig-4}(j).
This should be expected, to the extent that these diagrams contain self-energy insertions of the fermionic type (in diagrams 4(g)-4(h)) or of the bosonic type
(in diagrams 4(i)-4(j)).

By a related argument, one would expect on physical grounds the decoration at the lowest order in the disorder to be attributed directly to a self-energy of the bosonic type and not to diagrams \ref{fig-4}(g)-\ref{fig-4}(j) as a whole, in analogy to what was done in Fig.\ref{fig-2}(a) for the self-energy of the fermionic type.
This implies that these self-energy insertions of the bosonic type should be re-summed to infinite order, so as to replace the bare ladder $\Gamma_{0}$ by
a modified ladder $\Gamma$ suitably dressed by disorder.
This re-summation is convenient also from a mathematical point of view because diagrams \ref{fig-4}(g)-\ref{fig-4}(j) as they stand would diverge in the infrared upon approaching $T_{c}^{(0)}$ from above, since they contain two bare ladders $\Gamma_{0}$ with the same wave vector and frequency. 

The numerical calculation of the above diagrams can be performed with moderate effort in the coupling range $- 2 \lesssim (k_{F} a_{F})^{-1} \lesssim +2$.
It is then relevant to complement the numerical calculation of the above diagrams with analytic estimates which can be separately provided in the (extreme) BCS and BEC limits, that lie outside the above range where specific simplifying approximations hold.
These analytic estimates are discussed in detail in the two next sub-sections.

\vspace{0.1cm}
\begin{center}
{\bf C. Analytic estimates in the strong-coupling (BEC) limit}
\end{center}
\vspace{0.1cm}

In the BEC limit where $(k_{F} a_{F})^{-1} \gg 1$, the fermionic chemical potential $\mu$ is the largest energy scale in the problem and approaches asymptotically the value $- \varepsilon_{0}/2$ where $\varepsilon_{0}=(m a_{F})^{-1}$ is the binding energy of the two-fermion problem in vacuum.
In the present context, we further assume that $|\mu| \gg p_{0}^{2}/(2m)$, which corresponds to consider the size of the composite bosons smaller that the typical correlation length $\sim p_{0}^{-1}$ of the disorder \cite{footnote-xi_pair}.

\begin{figure}[h]
\includegraphics[angle=0,width=8.2cm]{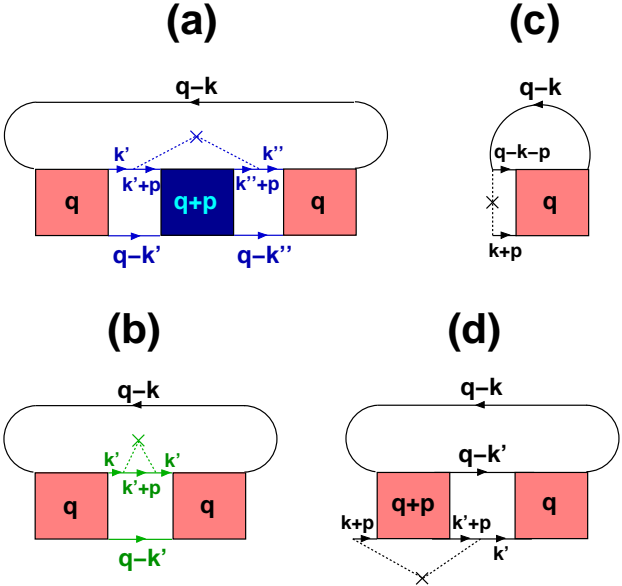}
\caption{The four diagrams here reported correspond, in the order, to diagrams of Figs.\ref{fig-4}(i), \ref{fig-4}(h), \ref{fig-4}(c), and \ref{fig-4}(d), where 
              the internal wave vectors and frequencies are indicated in four-vector notation $k=(\mathbf{k},\omega_{n})$ and $q=(\mathbf{q},\Omega_{\nu})$. 
              In addition, $\mathbf{p}$ is the wave vector associated with the impurity potential (\ref{impurity-potential}).}
\label{fig-5}
\end{figure}

Out of the diagrams drawn in Fig.\ref{fig-4}, we explicitly consider the BEC limit of diagrams \ref{fig-4}(i)-\ref{fig-4}(j) since they turn out to be the most important ones in the limit, but also of diagrams \ref{fig-4}(h) since they contain a kind of bosonic self-energy insertion to be compared with those of diagrams \ref{fig-4}(i)-\ref{fig-4}(j).
(The analytic estimate of diagram \ref{fig-4}(g) is not reported here explicitly since it turns out to be equivalent to that of diagrams \ref{fig-4}(h).)

In addition, we shall consider the BEC limit of diagrams \ref{fig-4}(c) which will be retained in the final numerical calculations owing to their relevance in the BCS limit, as well as 
of diagrams \ref{fig-4}(d)-\ref{fig-4}(e) in order to evidence their fermionic character.
(The analytic estimate of diagrams 4(a) and 4(b) turns out to be equivalent to that of diagrams 4(c), and the analytic estimate of diagrams \ref{fig-4}(f) turns out to be equivalent to that of diagrams \ref{fig-4}(d)-\ref{fig-4}(e). Accordingly, they will not be reported here explicitly.)

The above selected diagrams are redrawn for convenience in Fig.\ref{fig-5}, where the internal wave vectors and frequencies are also reported in order to identify and manipulate their analytic expressions.

In the BEC limit, the bare ladder propagator $\Gamma_{0}(\mathbf{q},\Omega_{\nu})$ of Fig.\ref{fig-2}(c) with wave vector $\mathbf{q}$ and bosonic Matsubara frequency 
$\Omega_{\nu} = 2 \pi k_{B} T \nu$ ($\nu$ integer) acquires a polar structure similar to that of a free-boson propagator \cite{Haussmann-93},\cite{Strinati-12}:
\begin{equation}
\Gamma_{0}(\mathbf{q},\Omega_{\nu}) \, \simeq \, - \left( \frac{8\pi}{m^{2} a_{F}} \right) \, \frac{1}{i\Omega_{\nu} \, - \, 
\mathbf{q}^{2}/(4m) \, + \, \mu_{B}}                                                                                                
\label{bare-bosonic-propagator}
\end{equation}
\noindent
with bosonic chemical potential $\mu_{B} = 2\mu+\varepsilon_{0}$.
The form (\ref{bare-bosonic-propagator}) excludes the high-energy contribution of a cut in the complex frequency plane, which is important to account for the scattering processes between two composite bosons made up of fermions pairs in vacuum \cite{KC-2005} but is irrelevant for the diagrams we consider in the present context (a feature that we have also verified numerically).
We are therefore going to use the asymptotic form (\ref{bare-bosonic-propagator}) of $\Gamma_{0}$ in the following estimates.

Since the four diagrams of Figs.\ref{fig-4}(i)-\ref{fig-4}(j) are all numerically equal, we have redrawn only one of them in Fig.\ref{fig-5}(a).
Upon extracting the bosonic self-energy insertions (that are evidenced in the central part of the diagrams), we obtain for their overall analytic contribution:
\begin{eqnarray}
& & \Sigma_{4i-4j}^{B}(\mathbf{q},\Omega_{\nu}) = 4 \gamma \int \! \frac{d\mathbf{k'}}{(2 \pi)^{3}} k_{B} T \sum_{n'} \int \! \frac{d\mathbf{k''}}{(2 \pi)^{3}} 
k_{B} T \sum_{n''}                                                                                                                                                                                               \nonumber \\
& \times & G_{0}(\mathbf{k'},\omega_{n'}) \, G_{0}(\mathbf{k''},\omega_{n''}) \, G_{0}(\mathbf{q}-\mathbf{k'},\Omega_{\nu}-\omega_{n'})         \nonumber \\ 
& \times & G_{0}(\mathbf{q}-\mathbf{k''},\Omega_{\nu}-\omega_{n''}) \, \int_{|\mathbf{p}|<p_{0}} \! \frac{d\mathbf{p}}{(2 \pi)^{3}} \,
G_{0}(\mathbf{k'}+\mathbf{p},\omega_{n'})                                                                                                                                                        \nonumber \\
& \times & G_{0}(\mathbf{k''+\mathbf{p}},\omega_{n''}) \,\, \Gamma_{0}(\mathbf{q}+\mathbf{p},\Omega_{\nu})
\label{bosonic-self-energy-4i-j}
\end{eqnarray}
\noindent
where $G_{0}(\mathbf{k},\omega_{n}) = (i\omega_{n} - \xi(\mathbf{k}))^{-1}$ is the bare fermionic propagator.
By our assumptions, we consider not only $|\mu| \gg p_{0}^{2}/(2m)$ but also $|\mu| \gg \mathbf{q}^{2}/(2m)$ and $|\mu| \gg |\Omega_{\nu}|$ for the bosonic energy scales of relevance.
We can thus neglect the wave vector $\mathbf{p}$ where it appears in a pair of $G_{0}$, as well as $\mathbf{q}$ and $\Omega_{\nu}$ where they appear
in a different pair of $G_{0}$.
The expression (\ref{bosonic-self-energy-4i-j}) then reduces to:
\begin{eqnarray}
& & \Sigma_{4i-4j}^{B}(\mathbf{q},\Omega_{\nu}) \simeq 4 \gamma 
\int_{|\mathbf{p}|<p_{0}} \! \frac{d\mathbf{p}}{(2 \pi)^{3}} \, \Gamma_{0}(\mathbf{q}+\mathbf{p},\Omega_{\nu})         \nonumber \\ 
& \times & \left( \! \int \! \frac{d\mathbf{k}}{(2 \pi)^{3}} k_{B} T \sum_{n} G_{0}(\mathbf{k},\omega_{n})^{2} \, G_{0}(-\mathbf{k},-\omega_{n}) \! \right)^{2} \! .                                                                                                                                                                                                                                                                                                                                                                                                              
\label{bosonic-self-energy-4i-j-approximate}
\end{eqnarray}
\noindent
With the result 
\begin{equation}
\int \! \frac{d\mathbf{k}}{(2 \pi)^{3}} k_{B} T \sum_{n} G_{0}(\mathbf{k},\omega_{n})^{2} G_{0}(-\mathbf{k},-\omega_{n}) \simeq - \frac{m^{2} a_{F}}{8 \pi}
\label{approximate-integral-BEC_limit}
\end{equation}
\noindent
that holds in the BEC limit (cf., e.g., Ref.\cite{Strinati-12}), we get eventually:
\begin{eqnarray}
& & - \left( \frac{8\pi}{m^{2} a_{F}} \right) \Sigma_{4i-4j}^{B}(\mathbf{q},\Omega_{\nu})            \label{bosonic-self-energy-4i-j-approximate-final} \\
& \simeq & 4 \gamma \, \int_{|\mathbf{p}|<p_{0}} \! \frac{d\mathbf{p}}{(2 \pi)^{3}} \, \frac{1}{i\Omega_{\nu} - (\mathbf{p}+\mathbf{q})^{2}/(4m) + \mu_{B}} \, .
\nonumber                                                                                               
\end{eqnarray}

The right-hand side of this expression represents the simplest self-energy process for non-interacting composite-like bosons of mass $2m$ subject to the impurity potential (\ref{impurity-potential}), and is the analogue of the expression (\ref{simplest-self-energy}) for non-interacting fermions subject to the same impurity potential.
Keeping here also finite values of the frequency $\Omega_{\nu}$, we  obtain:
\begin{eqnarray}
& & \int_{|\mathbf{p}|<p_{0}} \! \frac{d\mathbf{p}}{(2 \pi)^{3}} \, \frac{1}{i\Omega_{\nu} - (\mathbf{p}+\mathbf{q})^{2}/(4m) + \mu_{B}}                       
\nonumber \\
& \simeq & - \frac{2m p_{0}}{\pi^{2}} \, - \, i \, \frac{(4m)^{3/2}}{4 \pi} \, \mathrm{Re} \{ \sqrt{\mu_{B} + i \Omega_{\nu}} \} \, \mathrm{sgn}(\Omega_{\nu})
\nonumber \\ 
& = & - \frac{2m p_{0}}{\pi^{2}} \, + \, i \, \frac{(4m)^{3/2}}{4 \pi} \, \mathrm{Im} \{ \sqrt{- \mu_{B} - i \Omega_{\nu}} \} \, .    
\label{self-energy-point-like-bosons}                                                                                       
\end{eqnarray}
\noindent
Note that, through the identity that we have used in the last line, the result obtained in Ref.\cite{Vinokur-2002} in the limit of non-interacting point bosons in the normal phase is recovered, to the extent that $|\mathrm{Re} \{ \sqrt{- \mu_{B} - i \Omega_{\nu}} \}| \ll p_{0}/(\pi m^{1/2})$ for the relevant range of frequencies.

Note further that the factor $4$ on the right-hand side of Eq.(\ref{bosonic-self-energy-4i-j-approximate-final}) (which originates from the four distinct diagrams of Figs.\ref{fig-4}(i)-\ref{fig-4}(j)) consistently accounts for the presence of two fermions within a composite boson and for the fact that the impurity potential enters squared.
The factor $-8\pi/(m^{2} a_{F})$ on the left-hand side of Eq.(\ref{bosonic-self-energy-4i-j-approximate-final}) is instead required by the mapping from the original fermionic diagrammatic structure to the effective bosonic diagrammatic structure, as it is also evident from Eq.(\ref{bare-bosonic-propagator}).

\emph{In this way, we have identified the main features of the BCS-BEC crossover in the presence of a weak impurity disorder, following the evolution from non-interacting fermions in the extreme BCS limit to non-interacting composite bosons in the extreme BEC limit}.

Continuing through our program, we consider (one of) the diagrams of Fig.\ref{fig-4}(h), from which we extract the bosonic-like self-energy insertion $\Sigma_{4h}^{B}$ appearing in the middle of Fig.\ref{fig-5}(b).
Its analytic expression reads:
\begin{eqnarray}
\Sigma_{4h}^{B}(\mathbf{q},\Omega_{\nu}) & = & \int \! \frac{d\mathbf{k}}{(2 \pi)^{3}} \, k_{B} T \sum_{n} \, G_{0}(\mathbf{k},\omega_{n})^{2}    \nonumber \\
& \times & G_{0}(\mathbf{q}-\mathbf{k},\Omega_{\nu}-\omega_{n}) \, \Sigma_{2a}(\mathbf{k},\omega_{n}) 
\label{bosonic-self-energy-4h}
\end{eqnarray}
\noindent
where $\Sigma_{2a}$ is the fermionic self-energy of Fig.\ref{fig-2}(a).
Neglecting $\mathbf{q}$ and $\Omega_{\nu}$ in the argument of $G_{0}$, using again the approximate result (\ref{approximate-integral-BEC_limit}), and recalling the expression (\ref{approximate-self-energy-BEC}) for $\Sigma_{2a}$ in the BEC limit, we obtain:
\begin{eqnarray}
- \left( \frac{8\pi}{m^{2} a_{F}} \right) \Sigma_{4h}^{B}(\mathbf{q},\Omega_{\nu}) & \simeq & - \frac{\gamma \, p_{0}^{3}}{6\pi^{2} |\mu|}   
\label{bosonic-self-energy-4h-approximate-final} \\
& \simeq & - \frac{m \gamma p_{0}}{3 \pi^{2}} \left( \frac{p_{0}}{k_{F}} \right)^{2} (k_{F} a_{F})^{2}     \nonumber             
\end{eqnarray}
\noindent
which is negligible with respect to the results (\ref{bosonic-self-energy-4i-j-approximate-final}) and (\ref{self-energy-point-like-bosons}) under the assumption that 
$|\mu| \gg p_{0}^{2}/(2m)$.

We next consider (one of) the diagrams of Fig.\ref{fig-4}(c) as redrawn in Fig.\ref{fig-5}(c).
Its analytic expression reads:
\begin{eqnarray}
& & \Sigma_{4c}(\mathbf{k},\omega_{n}) = - \gamma \, \int \! \frac{d\mathbf{q}}{(2 \pi)^{3}} \, k_{B} T \, \sum_{\nu}                             \nonumber \\ 
& \times & \Gamma_{0}(\mathbf{q},\Omega_{\nu}) \,\, G_{0}(\mathbf{q}-\mathbf{k},\Omega_{\nu}-\omega_{n})                                       \label{self-energy-4c} \\ 
& \times & \int_{|\mathbf{p}|<p_{0}} \! \frac{d\mathbf{p}}{(2 \pi)^{3}} \,          
G_{0}(\mathbf{k}+\mathbf{p},\omega_{n}) \, G_{0}(\mathbf{q}-\mathbf{k}-\mathbf{p},\Omega_{\nu}-\omega_{n}) \, .                            \nonumber                                                                                                                                                                                                                                                                                                                                                                                                             
\end{eqnarray}
\noindent
Neglecting $\mathbf{p}$ as well as $\mathbf{q}$ and $\Omega_{\nu}$ with respect to $|\mu|$ in all the $G_{0}$, we obtain approximately ($\eta$ being a positive infinitesimal):

\begin{eqnarray}
& & \Sigma_{4c}(\mathbf{k},\omega_{n}) \simeq - \gamma \,\, G_{0}(\mathbf{k},\omega_{n}) \, G_{0}(-\mathbf{k},-\omega_{n})^{2}      \nonumber \\ 
& \times & \int \! \frac{d\mathbf{q}}{(2 \pi)^{3}} \, k_{B} T \, \sum_{\nu} \, \Gamma_{0}(\mathbf{q},\Omega_{\nu}) \, e^{i \Omega_{\nu} \eta} 
\int_{|\mathbf{p}|<p_{0}} \! \frac{d\mathbf{p}}{(2 \pi)^{3}}                                                                                                                          \nonumber \\                     
& \simeq & \frac{16}{9 \pi^{3}} \,\, (m \gamma p_{0}) \, \left( \frac{p_{0}}{k_{F}} \right)^{2} \, (k_{F} a_{F})^{5} 
\label{self-energy-4c-approximate}                                                                                                                                                                                                                                                                                                                                                                                                           
\end{eqnarray}
\noindent
which is strongly suppressed even with respect to the result (\ref{approximate-self-energy-BEC}) in the BEC limit where $|\mu|=(2 m a_{F}^{2})^{-1}$.
To obtain the last line of Eq.(\ref{self-energy-4c-approximate}) we have made use of the approximate expression (\ref{bare-bosonic-propagator}) and of the result
\begin{equation}
n_{B} = - \int \! \frac{d\mathbf{q}}{(2 \pi)^{3}} \, k_{B} T \, \sum_{\nu} \, \frac{e^{i \Omega_{\nu} \eta}}{i\Omega_{\nu} \, - \, \mathbf{q}^{2}/(4m) \, + \, \mu_{B}}                                         
\end{equation}
\noindent
which represents the density of composite bosons such that $n_{B}=n/2$.

Finally, it is relevant to estimate the behavior in the BEC limit of (one of) the diagrams \ref{fig-4}(d) as redrawn in Fig.\ref{fig-5}(d).
We obtain for temperatures of order $T_{c}$:
\begin{eqnarray}
& & \Sigma_{4d}(\mathbf{k},\omega_{n}) = - \gamma \, \int \! \frac{d\mathbf{k'}}{(2 \pi)^{3}} k_{B} T \sum_{n'} \int \! \frac{d\mathbf{q}}{(2 \pi)^{3}} k_{B} T \sum_{\nu}                             \nonumber \\ 
& \times & \int_{|\mathbf{p}|<p_{0}} \! \frac{d\mathbf{p}}{(2 \pi)^{3}} \, G_{0}(\mathbf{k}+\mathbf{p},\omega_{n}) \, G_{0}(\mathbf{k'}+\mathbf{p},\omega_{n'})  
\nonumber \\
& \times & G_{0}(\mathbf{k'},\omega_{n'}) G_{0}(\mathbf{q}-\mathbf{k'},\Omega_{\nu}-\omega_{n'}) G_{0}(\mathbf{q}-\mathbf{k},\Omega_{\nu}-\omega_{n})
\nonumber \\
& \times & \Gamma_{0}(\mathbf{q},\Omega_{\nu}) \,\,  \Gamma_{0}(\mathbf{q}+\mathbf{p},\Omega_{\nu}) 
\nonumber \\
& \simeq & - \, \frac{\gamma}{\mu^{2}} \, \left( \frac{8 \pi}{m^{2} \, a_{F}} \right) \,  \int_{|\mathbf{p}|<p_{0}} \! \frac{d\mathbf{p}}{(2 \pi)^{3}} \,
\int \! \frac{d\mathbf{q}}{(2 \pi)^{3}}
\label{self-energy-4d}   \\ 
& \times & \frac{ [ f_{B}(\xi_{B}(\mathbf{q})) - f_{B}(\xi_{B}(\mathbf{q}+\mathbf{p}))]}{\xi_{B}(\mathbf{q}) - \xi_{B}(\mathbf{q}+\mathbf{p})}
\simeq  \frac{64}{3 \pi^{3}} \, (m \gamma p_{0}) \, (k_{F} a_{F})^{3}        
\nonumber                                                                                                                                                                                                                                                                                                                                                                                                                                        
\end{eqnarray}
\noindent
where $f_{B}(\epsilon) = (e^{\epsilon/(k_{B}T)} - 1)^{-1}$ is the Bose function and $\xi_{B}(\mathbf{q}) = \mathbf{q}^{2}/(4m) - \mu_{B}$.
This is also strongly suppressed with respect to the result (\ref{approximate-self-energy-BEC}) in the BEC limit.

The above estimates hold in the \emph{extreme\/} BEC limit and require that $|\mu|/E_{F} \gg (p_{0}/k_{F})^{2}$, a condition which is quite difficult to satisfy in practice numerically. 
When $(k_{F} a_{F})^{-1} = + 2$, for instance, $|\mu|/E_{F} \approx 4$ while $(p_{0}/k_{F})^{2} \approx 25$ for the smallest value $p_{0}/k_{F} \approx 5$ that we can use for the numerical results of the critical temperature to be (essentially) independent of $p_{0}$.
For the purpose of testing our numerical codes that span the BCS-BEC crossover against the analytic results
(\ref{bosonic-self-energy-4i-j-approximate-final})-(\ref{self-energy-point-like-bosons}), (\ref{bosonic-self-energy-4h-approximate-final}),
(\ref{self-energy-4c-approximate}), and (\ref{self-energy-4d}) that are available in the BEC limit, we have thus made a number of specific runs of the numerical codes even up to $(k_{F} a_{F})^{-1} \approx +50$, in order to get agreement within a few per cent between the above analytic estimates and the numerical results.

\vspace{0.1cm}
\begin{center}
{\bf D. Analytic estimates in the weak-coupling (BCS) limit}
\end{center}
\vspace{0.1cm}

In the BCS limit where $(k_{F} a_{F})^{-1} \ll -1$, the Fermi energy is the largest energy scale in the problem and the bare ladder propagator $\Gamma_{0}$ can be approximated by the constant value $(- 4 \pi a_{F}/m)$ (except for an irrelevant narrow temperature window about $T_{c}^{(0)}$).
One then expects only diagrams \ref{fig-4}(a)-4(c) with the smallest number of $\Gamma_{0}$ to mostly contribute in this limit.

\begin{figure}[h]
\includegraphics[angle=0,width=8.5cm]{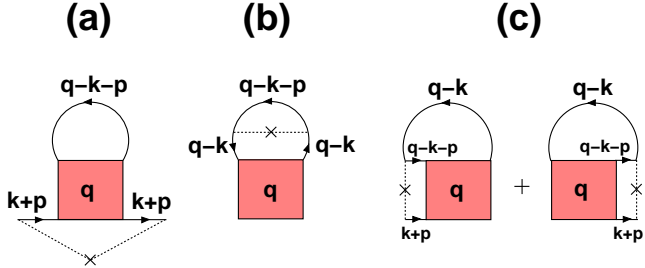}
\caption{The three diagrams here reported correspond, in the order, to the diagrams of Figs.\ref{fig-4}(a), \ref{fig-4}(b), and \ref{fig-4}(c), 
              where now the internal wave vectors and frequencies are also indicated with the four-vector notation of Fig.\ref{fig-5}.}
\label{fig-6}
\end{figure}

To obtain analytic estimates of these diagrams, we redraw them in Fig.\ref{fig-6} where the wave vectors and frequencies of the single-particle propagators $G_{0}$ are explicitly indicated. 
Approximating $\Gamma_{0}$ by a constant considerably simplifies these estimates as shown in the following.

For the diagram Fig.\ref{fig-6}(a) we obtain: 
\begin{eqnarray}
& & \Sigma_{4a}(\mathbf{k},\omega_{n}) \simeq \gamma \left( \frac{4 \pi a_{F}}{m} \right) \int \! \frac{d\mathbf{q}}{(2 \pi)^{3}} \, k_{B} T \, \sum_{\nu}  \nonumber \\  
& \times & \int_{|\mathbf{p}|<p_{0}} \! \frac{d\mathbf{p}}{(2 \pi)^{3}} G_{0}(\mathbf{k}+\mathbf{p},\omega_{n})^{2}
G_{0}(\mathbf{q}-\mathbf{k}-\mathbf{p},\Omega_{\nu}-\omega_{n}) \nonumber \\
& = & - \left( \! \frac{4 \pi a_{F}}{m} \! \right) \frac{n}{2} \frac{\partial}{\partial \mu} \Sigma_{2a}(\mathbf{k},\omega_{n}) \nonumber \\
& \simeq &  \frac{2 \, i}{3 \, \pi} \, (k_{F} a_{F}) \mathrm{Im} \{ \Sigma_{2a}(\mathbf{k},\omega_{n}) \}
\label{self-energy-4a-approximate-BCS}                                                                                                                                                                                                                                                                                                                                                                                                          
\end{eqnarray}
\noindent
where $\mathrm{Im} \{ \Sigma_{2a}(\mathbf{k},\omega_{n}) \}$ can be read off from the right-hand side of Eq.(\ref{approximate-self-energy-BCS}).

For the diagram Fig.\ref{fig-6}(b) we obtain instead:
\begin{eqnarray}
& & \Sigma_{4b}(\mathbf{k},\omega_{n}) \simeq \gamma \left( \frac{4 \pi a_{F}}{m} \right) \int \! \frac{d\mathbf{q}}{(2 \pi)^{3}} \, k_{B} T \, \sum_{\nu}  \nonumber \\  
& \times & \! G_{0}(\mathbf{q}-\mathbf{k},\Omega_{\nu}-\omega_{n})^{2} \!\!\! 
\int_{|\mathbf{p}|<p_{0}} \!\! \frac{d\mathbf{p}}{(2 \pi)^{3}} G_{0}(\mathbf{q}-\mathbf{k}-\mathbf{p},\Omega_{\nu}-\omega_{n})
\nonumber \\
& = & \left( \frac{4 \pi a_{F}}{m} \right) \int \!\! \frac{d\mathbf{k'}}{(2 \pi)^{3}} \, k_{B} T \, \sum_{n'} G_{0}(\mathbf{k'},\omega_{n'})^{2} \,
\Sigma_{2a}(\mathbf{k'},\omega_{n'})
\nonumber \\
& \simeq & \! \left( \! \frac{4 \pi a_{F}}{m} \! \right) \! \chi_{0}^{(\mathrm{ph})}(0,0) \Sigma_{2a}(0,0) = - \frac{2}{\pi} (k_{F} a_{F}) \Sigma_{2a}(0,0)
\label{self-energy-4b-approximate-BCS}                                                                                                                                                                                                                                                                                                                                                                                                           
\end{eqnarray}
\noindent
where $\chi_{0}^{(\mathrm{ph})}(0,0) = - N_{0}$ is the value of the polarization (Lindhard) function of zero arguments per spin component at low temperature.

Finally, for (one of) the diagrams Fig.\ref{fig-6}(c) we obtain:
\begin{eqnarray}
& & \Sigma_{4c}(\mathbf{k},\omega_{n}) \simeq \gamma \left( \frac{4 \pi a_{F}}{m} \right) \int \! \frac{d\mathbf{q}}{(2 \pi)^{3}} \, k_{B} T \, \sum_{\nu}  \nonumber \\  
& \times & G_{0}(\mathbf{q}-\mathbf{k},\Omega_{\nu}-\omega_{n})
\int_{|\mathbf{p}|<p_{0}} \! \frac{d\mathbf{p}}{(2 \pi)^{3}} G_{0}(\mathbf{k}+\mathbf{p},\omega_{n})
\nonumber \\  
& \times & G_{0}(\mathbf{q}-\mathbf{k}-\mathbf{p},\Omega_{\nu}-\omega_{n})
\nonumber \\
& = & \gamma \left( \frac{4 \pi a_{F}}{m} \right) \int_{|\mathbf{p}|<p_{0}} \! \frac{d\mathbf{p}}{(2 \pi)^{3}} \,\, \chi_{0}^{(\mathrm{ph})}(\mathbf{p},0) \, 
G_{0}(\mathbf{k}+\mathbf{p},\omega_{n})
\nonumber \\
& \approx & \left( \frac{4 \pi a_{F}}{m} \right) \, \chi_{0}^{(\mathrm{ph})}(0,0) \, \Sigma_{2a}(0,0) 
\label{self-energy-4c-approximate-BCS}                                                                                                                                                                                                                                                                                                                                                                                                           
\end{eqnarray}
\noindent
which essentially coincides with the result (\ref{self-energy-4b-approximate-BCS}).

The expressions (\ref{self-energy-4a-approximate-BCS})-(\ref{self-energy-4c-approximate-BCS}) show that the self-energies of Figs.\ref{fig-4}(a)-\ref{fig-4}(c)
are all much smaller than the self-energy of Fig.\ref{fig-2}(a) in the \emph{extreme\/} BCS limit whereby $k_{F} |a_{F}| \ll 1$.
In practice, the numerical calculation of these diagrams can be extended with reasonable effort down to $(k_{F} a_{F})^{-1} \approx -2$, where their contributions may turn out to be more relevant than what expected from the above estimates.
However, again for the purpose of testing our numerical codes that span the BCS-BEC crossover against the analytic results
(\ref{self-energy-4a-approximate-BCS}), (\ref{self-energy-4b-approximate-BCS}), and (\ref{self-energy-4c-approximate-BCS}) available in the BCS limit, we have also made a number of runs of the numerical codes down to $(k_{F} a_{F})^{-1} = -5.0$ using again the value $p_{0}/k_{F} = 5$, obtaining in all cases a few per cent agreement between the numerical results and the above theoretical estimates.

\section{III. Numerical results}
\label{sec:numerical_results}

In the previous Section, we have identified the fermionic self-energy diagrams that include the effect of disorder at the lowest significant order, over and above the t-matrix approximation spanning the BCS-BEC crossover.
We have also obtained analytic estimates for these diagrams separately in the BCS and BEC limits where such estimates are possible, and checked the analytic results against accurate numerical calculations.

We pass now to extend the numerical calculations to the whole BCS-BEC crossover, aiming at establishing a hierarchy on the relative importance of
the various diagrams.
This will eventually enable us to select a limited subset of diagrams, that will be retained in the final calculation of physical quantities like the critical temperature and chemical potential across the BCS-BEC crossover in the presence of disorder.

\vspace{0.1cm}
\begin{center}
{\bf A. Numerical calculation of the self-energy diagrams in the presence of weak disorder throughout the BCS-BEC crossover}
\end{center}
\vspace{0.1cm}

We have calculated the wave-vector and frequency dependence of all the fermionic self-energy diagrams reported in Fig.\ref{fig-2} and  
Figs.\ref{fig-4}(a)-4(f), and of the bosonic-like self-energy insertions that enter the diagrams of Figs.\ref{fig-4}(g)-4(j), using the values of the critical temperature
$T_{c}^{(0)}$ and of the corresponding chemical potential $\mu^{(0)}(T_{c}^{(0)})$ of the clean system reported in Fig.\ref{fig-1} as functions of coupling.

\begin{figure}[h]
\includegraphics[angle=0,width=9.0cm]{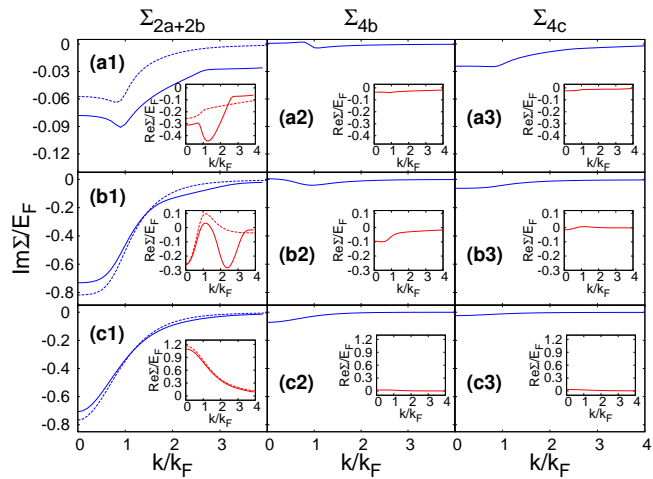}
\caption{Imaginary part of the fermionic self-energies $\Sigma_{2\mathrm{a}+2\mathrm{b}}$, $\Sigma_{4\mathrm{b}}$, and $\Sigma_{4\mathrm{c}}$ (in units of $E_{F}$) 
              vs the wave vector $k$ (in units of $k_{F}$) taken at $\omega_{n=0}=\pi k_{B} T_{c}^{(0)}$ with $T_{c}^{(0)}$ and $\mu^{(0)}(T_{c}^{(0)})$ of the clean system for
              three couplings: $(k_{F} a_{F})^{-1} = -1$ (top panels), $(k_{F} a_{F})^{-1} = 0$ (middle panels), $(k_{F} a_{F})^{-1} = +1$ (bottom panels). The disorder parameter 
              is $\tilde{\gamma} = 0.01$. The real part is given in the corresponding insets. In the left panels, full (dashed) lines refer to the presence (absence) of disorder.}
\label{fig-7}
\end{figure}

\begin{figure}[t]
\includegraphics[angle=0,width=9.0cm]{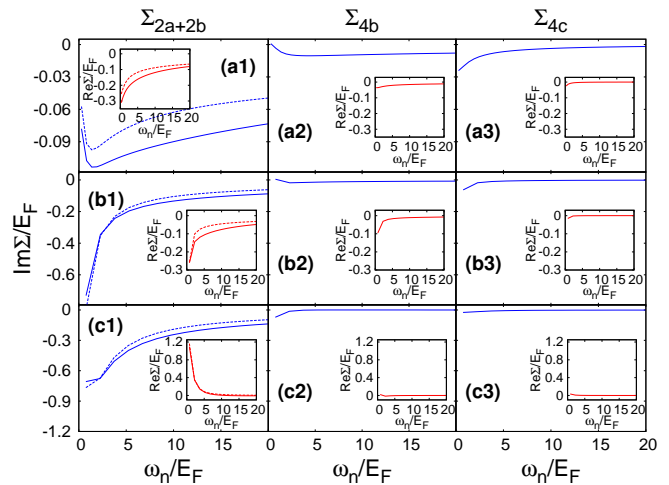}
\caption{Imaginary part of the fermionic self-energies $\Sigma_{2\mathrm{a}+2\mathrm{b}}$, $\Sigma_{4\mathrm{b}}$, and $\Sigma_{4\mathrm{c}}$ (in units of $E_{F}$) 
              vs the Matsubara frequency $\omega_{n}$ (in units of $E_{F}$) taken at $k=0$ with $T_{c}^{(0)}$ and $\mu^{(0)}(T_{c}^{(0)})$ of the clean system for the same 
              couplings of Fig.\ref{fig-7}. The disorder parameter is $\tilde{\gamma} = 0.01$. The real part is given in the corresponding insets. In the left panels, 
              full (dashed) lines refer to the presence (absence) of disorder.}
\label{fig-8}
\end{figure}

As an example, we show in Fig.\ref{fig-7} the dependence on the wave-vector of the fermionic self-energy $\Sigma_{2\mathrm{a}+2\mathrm{b}}$ (corresponding to the sum of the diagrams of Figs.\ref{fig-2}(a) and \ref{fig-2}(b)) and of the fermionic self-energies $\Sigma_{4\mathrm{b}}$ and $\Sigma_{4\mathrm{c}}$ (corresponding to the diagrams of Figs.\ref{fig-4}(b) and \ref{fig-4}(c), in the order) taken at the Matsubara frequency 
$\omega_{n=0}=\pi k_{B} T_{c}^{(0)}$, for three different couplings with the value $\tilde{\gamma} = 0.01$ for the dimensionless disorder parameter.
Each self-energy takes into account the multiplicity factor of the diagrams (for instance, two diagrams contribute to the self-energy $\Sigma_{4\mathrm{c}}$ as shown in 
Fig.\ref{fig-4}(c)).
The corresponding dependence on the Matsubara frequency is reported in Fig.\ref{fig-8} for vanishing wave vector.

\begin{figure}[h]
\includegraphics[angle=0,width=8.5cm]{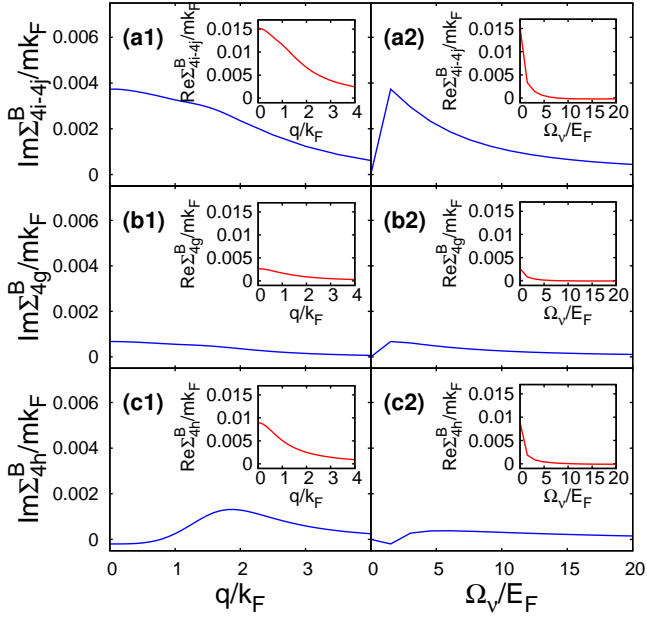}
\caption{Imaginary part (and real part in the insets) of the bosonic-like self-energies $\Sigma_{4\mathrm{i}-4\mathrm{j}}^{B}$, $\Sigma_{4\mathrm{g}}^{B}$, and 
$\Sigma_{4\mathrm{h}}^{B}$ (in units of $m k_{F}$) vs the wave vector $q$ (in units of $k_{F}$) taken at $\Omega_{\nu} =2 \pi T_{c}^{(0)}$ for the imaginary part and at $\Omega_{\nu} =0$ for the real part (left panels), and vs the Matsubara frequency $\Omega_{\nu}$ (in units of $E_{F}$) taken at $q=0$ (right panels). The coupling is $(k_{F} a_{F})^{-1}=0$ and the disorder parameter is $\tilde{\gamma} = 0.01$.}
\label{fig-9}
\end{figure}

A similar analysis can be performed for the bosonic-like self-energy insertions that enter the diagrams of Figs.\ref{fig-4}(g)-4(j).
This is shown in Fig.\ref{fig-9} where the wave-vector and frequency dependence of diagrams \ref{fig-4}(i)-4(j) is compared with that of diagrams
\ref{fig-4}(g) and \ref{fig-4}(h).
From this comparison we conclude that diagrams \ref{fig-4}(g) and \ref{fig-4}(h) can be neglected with respect to diagrams \ref{fig-4}(i)-4(j).

Accordingly, in Figs.\ref{fig-7} and \ref{fig-8} the ladder propagator entering the diagram of Fig.\ref{fig-2}(b) has been dressed only with the bosonic-like self-energy insertions of Figs.\ref{fig-4}(i)-4(j), thus forming the dressed ladder propagator $\Gamma$ defined by Eq.(\ref{dressed-ladder-propagator}) below, which in the presence of disorder replaces the bare ladder propagator $\Gamma_{0}$.
Correspondingly, in the left panels of Figs.\ref{fig-7} and \ref{fig-8} and in the associated insets, full and dashed lines refer to the presence and the absence of disorder, respectively.

A general conclusion that can be drawn from the above analysis is that, when a given self-energy is dominant with respect to another one at (or near) zero wave vector and frequency, it also remains dominant for all wave vectors and frequencies.
For this reason, to assess (at least in a preliminary way) the relative importance of the various self-energies one may look at their imaginary parts at zero wave vector and frequency, since it will be the imaginary part of a given self-energy to mostly affect the value of the critical temperature (see below).

This comparison is shown in Fig.\ref{fig-10} throughout the whole BCS-BEC crossover, where the behavior of \emph{all} fermionic diagrams drawn in Figs.\ref{fig-2} and 
\ref{fig-4} is reported. 
Besides the self-energy $\Sigma_{2\mathrm{a}+2\mathrm{b}}$ which is by far the most dominant one for all couplings, at this level special consideration should apparently be given to the self-energies $\Sigma_{4\mathrm{b}}$ and $\Sigma_{4\mathrm{c}}$ (and possibly also to the self-energy $\Sigma_{4\mathrm{f}}$).

\begin{figure}[h]
\includegraphics[angle=0,width=8.3cm]{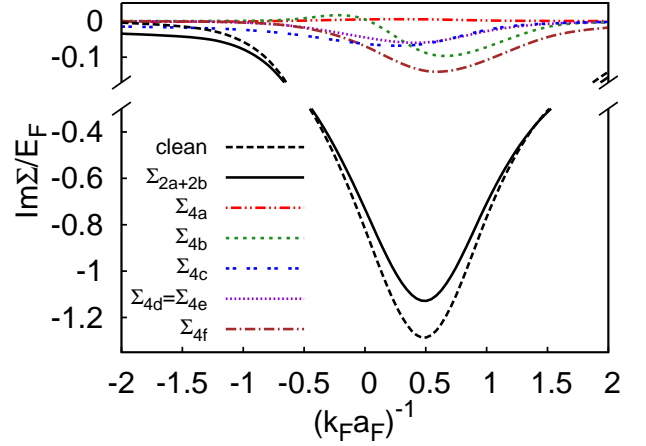}
\caption{Coupling dependence of  the imaginary part of the fermionic self-energies of Fig.\ref{fig-2} (both with the bare $\Gamma_{0}$ without disorder and with 
             the dressed $\Gamma$ with disorder, see the text) and of Fig.\ref{fig-4} (in units of $E_{F}$), taken at $k=0$ and $\omega_{n=0}=\pi k_{B} T_{c}^{(0)}$ 
             with $T_{c}^{(0)}$ and $\mu^{(0)}(T_{c}^{(0)})$ of the clean system. The disorder parameter is $\tilde{\gamma} = 0.01$.}
\label{fig-10}
\end{figure}

Here, the relevance of diagram $\Sigma_{4\mathrm{b}}$ can be expected by the self-consistent dressing by disorder of the upper line in the t-matrix self-energy $\Sigma_{2\mathrm{b}}$.
Consideration to the diagrams $\Sigma_{4\mathrm{c}}$ goes instead back to the theory of disordered interacting electrons in metals, where it is referred to as the Hartree correction to the self-energy \cite{LR-1985}.

A final assessment on the relative importance of the various fermionc self-energy diagrams is deferred to the calculation of the critical temperature in the presence of disorder throughout the BCS-BEC crossover, to be considered next.

\vspace{0.1cm}
\begin{center}
{\bf B. Numerical results for the critical temperature and chemical potential in the presence of weak disorder throughout the BCS-BEC crossover}
\end{center}
\vspace{0.1cm}

For a clean system, the critical temperature $T_{c}^{(0)}$ for the transition from the normal to the superfluid phase can be obtained in the grand-canonical ensemble in terms of the Thouless criterion $\Gamma_{0}^{-1}(\mathbf{q}=0,\Omega_{\nu}=0;\mu,T_{c}^{(0)}) = 0$ \cite{Schrieffer-1964}.
In the presence of disorder, this criterion has to be generalized to include scattering off the impurities.
For the present treatment of the BCS-BEC crossover, it is sufficient to consider the bosonic self-energy insertion of Figs.\ref{fig-4}(i)-4(j)
with $\Sigma_{4i-4j}^{B}(\mathbf{q},\Omega_{\nu})$ given by the expression (\ref{bosonic-self-energy-4i-j}) and obtain a modified value $T_{c}$ of the critical temperature from the condition:
\begin{equation}
\Gamma_{0}^{-1}(\mathbf{q}=0,\Omega_{\nu}=0;\mu,T_{c}) = \Sigma_{4i-4j}^{B}(\mathbf{q}=0,\Omega_{\nu}=0;\mu,T_{c})
\label{modified-Thouless-condition}
\end{equation}
\noindent
which is reminiscent of the Hugenholtz-Pines relation for point-like bosons \cite{HP-1959}.
More generally, we may define a dressed ladder propagator $\Gamma(\mathbf{q},\Omega_{\nu})$, such that
\begin{equation}
\Gamma^{-1}(\mathbf{q},\Omega_{\nu}) = \Gamma^{-1}_{0}(\mathbf{q},\Omega_{\nu}) - \Sigma_{4i-4j}^{B}(\mathbf{q},\Omega_{\nu}) \, ,
\label{dressed-ladder-propagator}
\end{equation}
\noindent
in terms of which the modified Thouless criterion (\ref{modified-Thouless-condition}) reads 
$\Gamma^{-1}(\mathbf{q}=0,\Omega_{\nu}=0;\mu,T_{c})=0$.

In addition, the chemical potential can be eliminated in favor of the particle density via the equation:
\begin{equation}
n \, = \, 2 \, \int \! \frac{d\mathbf{k}}{(2 \pi)^{3}} \, k_{B} T \, \sum_{n} \, e^{i \omega_{n} \eta} \, G(\mathbf{k},\omega_{n})
\label{density-equation}
\end{equation}
\noindent
where $G^{-1}(\mathbf{k},\omega_{n}) = G^{-1}_{0}(\mathbf{k},\omega_{n}) - \Sigma(\mathbf{k},\omega_{n})$ identifies a dressed fermionic propagator $G$ in terms of the bare $G_{0}$ and of the chosen fermionic self-energy $\Sigma$ \cite{footnote-replica_trick}.

From the analytic estimates and the numerical analysis that we have carried out for the various self-energy contributions reported in Figs.\ref{fig-2} and \ref{fig-4}
(as well from the further checks made on the effects of all self-energy contributions on the critical temperature, see below), \emph{we end up eventually in retaining}, besides the fermionic self-energy $\Sigma_{2\mathrm{a}+2\mathrm{b}}$ where the bare
$\Gamma_{0}$ is replaced by the dressed ladder $\Gamma$ of Eq.(\ref{dressed-ladder-propagator}) with the full dependence on $\mathbf{q}$ and 
$\Omega_{\nu}$, also the fermionic self-energies $\Sigma_{4\mathrm{b}}$ and $\Sigma_{4\mathrm{c}}$ where $\Gamma_{0}$ is again replaced by $\Gamma$ but now with the bosonic self-energy $\Sigma_{4i-4j}^{B}$ taken at $\mathbf{q}=0$ and $\Omega_{\nu}=0$.
Here, the inclusion of a limited degree of self-consistency originates from the need to avoid divergencies that may occur at the critical temperature when $\mathbf{q}=0$ and $\Omega_{\nu}=0$. 

\begin{figure}[t]
\includegraphics[angle=0,width=8.5cm]{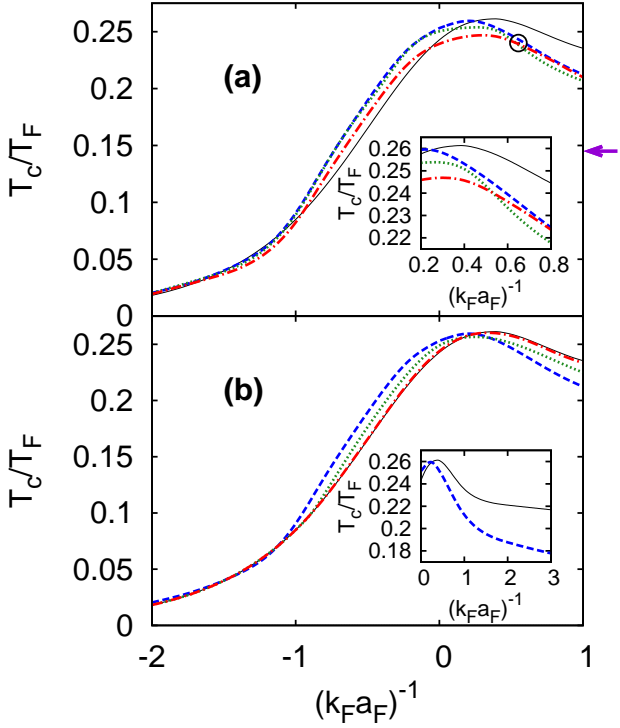}
\caption{Critical temperature $T_{c}$ (in units of $E_{F}$) vs the coupling parameter $(k_{F} a_{F})^{-1}$. In panel (a), the disorder parameter $\tilde{\gamma}$ is kept 
              at the value $0.01$ and various approximations for the self-energy are considered in the density equation: $\Sigma _{\mathrm{2a+2b}}$
              (dashed line), $\Sigma _{\mathrm{2a+2b+4c}}$ (dotted line), and $\Sigma _{\mathrm{2a+2b+4b+4c}}$ (dashed-dotted line). The result for the clean system is also 
              shown for comparison (full line). The circle evidences the region past the maximum of $T_{c}$ where in the presence of disorder all curves merge into a single one 
              (the inset makes this region more evident). The arrow on the right identifies the value of $T_{c}$ reached asymptotically in the BEC 
              limit in the presence of disorder.
              In panel (b), only the self-energy $\Sigma _{\mathrm{2a+2b}}$ is considered for various values of the disorder parameter $\tilde{\gamma}$:  $0$ 
              (full line), $0.001$ (dashed-dotted line), $ 0.005$ (dotted line), $ 0.01$ (broken line).
              The inset in panel (b) extends deeper to the BEC side of the crossover the curves of $T_{c}$ with $\tilde{\gamma} = 0$ (full line) and $\tilde{\gamma} = 0.01$   
              (dashed line).}
\label{fig-11}
\end{figure}

The results for the critical temperature $T_{c}$ throughout the BCS-BEC crossover obtained in this way in the presence of disorder are shown in Fig.\ref{fig-11}.
In panel (a) several approximations are considered for the fermionic self-energy (namely, $\Sigma _{\mathrm{2a+2b}}$, $\Sigma _{\mathrm{2a+2b+4c}}$, and $\Sigma _{\mathrm{2a+2b+4b+4c}}$) at fixed disorder, while in panel (b) only the main approximation $\Sigma _{\mathrm{2a+2b}}$ is retained for various degrees of disorder.
Note from panel (a) how the inclusion of $\Sigma _{\mathrm{4b}}$ results in a noticeable contribution to $T_{c}$ close to unitarity, slightly reducing the increase of $T_{c}$ obtained with $\Sigma _{\mathrm{2a+2b}}$ in the presence of disorder on the BCS side of unitarity.
In the next Section, we will discuss in more detail this increase of $T_{c}$ on the BCS side of unitarity and relate it to peculiar features of the single-particle density of states in the presence of interaction and disorder.
Note also how different approximations in the presence of disorder all give essentially the same result for $T_{c}$ past the coupling $(k_{F} a_{F})^{-1} \approx 0.5$, signaling the collapse of the underlying Fermi surface and the corresponding emergence of a predominantly bosonic character in the system.
We have also verified that the further inclusion of the fermionic self-energy $\Sigma _{\mathrm{4f}}$ (or else of the remaining fermionic self-energies out of those of Fig.\ref{fig-4}) does not change appreciably the above results. 

In the inset of Fig.\ref{fig-11}(b) the curves of $T_{c}$ obtained with the self-energy $\Sigma _{\mathrm{2a+2b}}$ with $\tilde{\gamma} = 0$ (full line) and $\tilde{\gamma} = 0.01$ (dashed line) are extended deeper to the BEC side of the crossover, to highlight the fact that \emph{the limiting value of the Bose-Einstein temperature $T_{\mathrm{BEC}}$ is much reduced by disorder}.
To this end, we have verified numerically that when $\tilde{\gamma} = 0.01$ this limiting value is approximately $0.14 \, T_{F}$ in agreement with the analytic estimate (\ref{Tc-BEC-with-disorder}) below, to be contrasted with the value $0.218 \, T_{F}$ of $T_{\mathrm{BEC}}$ for the clean system.

\begin{figure}[t]
\includegraphics[angle=0,width=8.5cm]{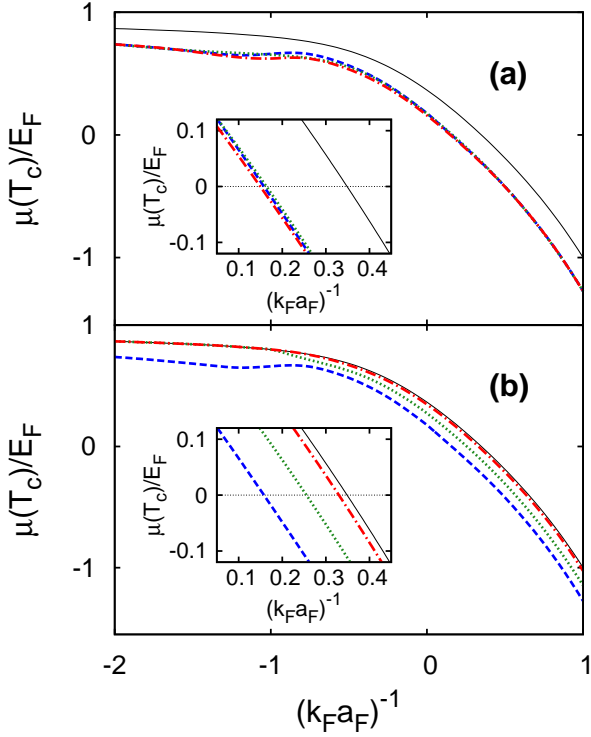}
\caption{Chemical potential $\mu(T_{c})$ at $T_{c}$ (in units of $E_{F}$) vs the coupling parameter $(k_{F} a_{F})^{-1}$. Conventions are the same of Fig.\ref{fig-11}.
             In panel (a) the disorder parameter $\tilde{\gamma}$ is kept fixed and various approximations for the self-energy are considered in the density equation, while 
             in panel (b) only the self-energy $\Sigma _{\mathrm{2a+2b}}$ is considered for various values of the disorder parameter.  In both panels, the insets amplify the 
             region where the chemical potential changes sign.}
\label{fig-12}
\end{figure}

The corresponding results for the chemical potential $\mu(T_{c})$ at $T_{c}$ are shown in Fig.\ref{fig-12}.
In contrast to $T_{c}$, for this quantity only minor differences result when considering various approximations for the fermionic self-energy.
Note how the presence of disorder makes the sign change of the chemical potential, which is a characteristic feature of the BCS-BEC crossover, to occur closer to unitarity with respect to the clean case.
We shall return to this point in the next Section while discussing the fate of the underlying remnant Fermi surface.

It is relevant to compare our numerical results for $T_{c}$ with the analytic estimates that can be obtained in the extreme BCS and BEC limits.
In particular, in the extreme BEC limit where the results (\ref{bosonic-self-energy-4i-j-approximate-final}) and (\ref{self-energy-point-like-bosons}) hold,
we obtain by expanding Eq.(\ref{density-equation}) in terms of the self-energy of Fig.\ref{fig-2}(b) with $\Gamma$ replacing $\Gamma_{0}$:

\begin{eqnarray}
\frac{n}{2} & \simeq & - \, \int \! \frac{d\mathbf{k}}{(2 \pi)^{3}} \, k_{B} T \, \sum_{n} \int \! \frac{d\mathbf{q}}{(2 \pi)^{3}} \, k_{B} T \, \sum_{\nu} 
\nonumber \\
& & \times \, G_{0}(\mathbf{k},\omega_{n})^{2} \, G_{0}(\mathbf{q}-\mathbf{k},\Omega_{\nu}-\omega_{n}) \,
\Gamma(\mathbf{q},\Omega_{\nu})
\nonumber \\
& \simeq & - \, \int \! \frac{d\mathbf{k}}{(2 \pi)^{3}} \, k_{B} T \, \sum_{n}  \, G_{0}(\mathbf{k},\omega_{n})^{2} \, G_{0}(-\mathbf{k},-\omega_{n})
\nonumber \\
& & \times \, \int \! \frac{d\mathbf{q}}{(2 \pi)^{3}} \, k_{B} T \, \sum_{\nu} \, e^{i \Omega_{\nu} \eta} \, \Gamma(\mathbf{q},\Omega_{\nu})
\nonumber \\
& \simeq & \left( \frac{m^{2} a_{F}}{8 \pi} \right) \, \int \! \frac{d\mathbf{q}}{(2 \pi)^{3}} \, k_{B} T \, \sum_{\nu} \, e^{i \Omega_{\nu} \eta} \,
\nonumber \\
& & \times \, \left[ \, \Gamma_{0}(\mathbf{q},\Omega_{\nu}) \, + \, \Gamma_{0}(\mathbf{q},\Omega_{\nu})^{2} \,\,
\Sigma_{4i-4j}^{B}(\mathbf{q},\Omega_{\nu}) \, \right]
\label{density-equation-BEC-limit-1}
\end{eqnarray}
\noindent
where the result (\ref{approximate-integral-BEC_limit}) has once more been used.
With the expressions (\ref{bare-bosonic-propagator}), (\ref{bosonic-self-energy-4i-j-approximate-final}), and (\ref{self-energy-point-like-bosons}) where $\mu_{B}$ is set to zero in order to identify $T_{c}$, Eq.(\ref{density-equation-BEC-limit-1}) becomes:
\begin{eqnarray}
\frac{n}{2} & \simeq & - \, \int \! \frac{d\mathbf{q}}{(2 \pi)^{3}} \, k_{B} T_{c} \, \sum_{\nu} \, e^{i \Omega_{\nu} \eta} \,
\frac{1}{i\Omega_{\nu} \, - \, \mathbf{q}^{2}/(4m)}  
\nonumber \\
& - & \frac{\gamma (4m)^{3/2}}{\pi} k_{B} T_{c} \sum_{\nu} e^{i \Omega_{\nu} \eta} \int \! \frac{d\mathbf{q}}{(2 \pi)^{3}} \,  
\frac{\sqrt{-i\Omega_{\nu}}}{\left( i\Omega_{\nu} - \frac{\mathbf{q}^{2}}{4m} \right)^{2}}
\nonumber \\
& = & \left( \frac{m k_{B} T_{c}}{\pi} \right)^{3/2} \zeta(3/2) + \frac{(4 \gamma) \, (2m)^{3}}{4 \, \pi^{2}} \, k_{B} T_{c}
\label{density-equation-BEC-limit-2}
\end{eqnarray}
\noindent
where $\zeta(3/2)$ is the Riemann zeta function of argument $3/2$.
With the mapping $n_{B}=n/2$, $m_{B}=2m$, and $\gamma_{B}=4\gamma$ between bosonic and fermionic quantities, 
Eq.(\ref{density-equation-BEC-limit-2}) recovers the expression obtained in Ref.\cite{Vinokur-2002} for non-interacting point-like bosons at the lowest order in the disorder.
Solving for $T_{c}$ by iteration, one gets accordingly from Eq.(\ref{density-equation-BEC-limit-2}):
\begin{equation}
T_{c} \simeq T_{c}^{(0)} \, \left( 1 \, - \, \frac{32 \, \gamma \, m^{3} \, k_{B} \, T_{c}^{(0)}}{3 \, \pi^{2} \, n} \right)
\label{Tc-BEC-with-disorder}
\end{equation}
\noindent
where $k_{B} T_{c}^{(0)}=(\pi/m) [n/(2 \zeta(3/2))]^{2/3}$ is here the Bose-Einstein temperature of the ideal gas.

In the extreme BCS limit, on the other hand, disorder affects the system only through the self-energy diagram of Fig.\ref{fig-2}(a).
Correspondingly, the critical temperature could be affected by dressing with this self-energy the fermion propagators $G_{0}$ in the rungs of the pair (ladder) propagator of Fig.\ref{fig-2}(c). 
One obtains for the elementary rung in the limit $\mathbf{q} \rightarrow 0$ and $\Omega_{\nu} \rightarrow 0$:
\begin{eqnarray}
& & \int \! \frac{d\mathbf{k}}{(2 \pi)^{3}} \, k_{B} T \, \sum_{n} \, G_{0}(\mathbf{k},\omega_{n}) \, G_{0}(\mathbf{q}-\mathbf{k},\Omega_{\nu}-\omega_{n})
\nonumber \\
& \longrightarrow & \, \int \! \frac{d\mathbf{k}}{(2 \pi)^{3}} \, k_{B} T \, \sum_{n} \, 
\frac{1}{ i \omega_{n} - \xi(\mathbf{k}) + \frac{i}{2 \tau} \, {\rm sgn}(\omega_{n}) }
\nonumber \\
& \times & \frac{1}{ i \Omega_{\nu} - i \omega_{n} - \xi(\mathbf{q}-\mathbf{k}) + \frac{i}{2 \tau} \, {\rm sgn}(\Omega_{\nu}-\omega_{n}) }
\nonumber \\
& & \overrightarrow{(\mathbf{q} = 0, \Omega_{\nu} = 0)} \,\, \int \! \frac{d\mathbf{k}}{(2 \pi)^{3}} \, \frac{1}{2 \, \xi_{\mathbf{k}}} \, \int_{- \infty}^{+ \infty} \! \frac{d \omega}{\pi} \, f_{F}(\omega) 
\nonumber \\
& \times & \frac{1}{2 \tau}  \, \left[ \frac{1}{(\omega + \xi(\mathbf{k}))^{2} + \frac{1}{(2 \tau)^{2}}} -  
\frac{1}{(\omega - \xi(\mathbf{k}))^{2} + \frac{1}{(2 \tau)^{2}}} \right]
\nonumber \\
& \simeq &  \int \! \frac{d\mathbf{k}}{(2 \pi)^{3}} \, \frac{1}{2 \, \xi(\mathbf{k})} \, \left[ f_{F}(-\xi(\mathbf{k})) - f_{F}(\xi(\mathbf{k})) \right]
\label{BCS-rung-with-disorder}
\end{eqnarray}
\noindent 
where $\tau^{-1}= 2 \pi N_{0} \gamma$ and $f_{F}(\epsilon) =(e^{\epsilon/(k_{B}T)} + 1)^{-1}$ is the Fermi function.
The last line of Eq.(\ref{BCS-rung-with-disorder}) depends on the assumption $(2 \tau)^{-1} \ll E_{F}$, which is met by our condition $m \gamma p_{0} \ll E_{F}$ with $p_{0} \gg k_{F}$ \cite{footnote-condition}.
This shows that the rung (\ref{BCS-rung-with-disorder}) (and therefore the critical temperature) is not affected by the presence of a weak disorder, in agreement with the Anderson theorem \cite{footnote-Anderson-theorem}. 

Finally, it is interesting to compare the contrasting results for the effect of a weak disorder on the critical temperature, which are obtained for a system of non-interacting fermions and non-interacting composite bosons, respectively, in the BCS and BEC limit, in the light of the apparent similarity between the expressions (\ref{simplest-self-energy}) and (\ref{bosonic-self-energy-4i-j-approximate-final}) for the relevant self-energies in the two cases.
The difference appears, in fact, in the resulting expressions (\ref{approximate-self-energy-BCS}) and (\ref{self-energy-point-like-bosons}), in the order, since to the Fermi statistics there corresponds a large value of $\mu$ in Eq.(\ref{approximate-self-energy-BCS}) while the Bose statistics allows 
$\mu_{B}$ to vanish in Eq.(\ref{self-energy-point-like-bosons}) at the critical temperature.
Accordingly, fermions are protected from the influence of disorder by the presence of a (large) Fermi surface which much limits the possible scattering processes by the impurities, while bosons are not.

\vspace{0.1cm}
\begin{center}
{\bf C. Effects of disorder on the Tan's contact}
\end{center}
\vspace{0.1cm}

It is further interesting to assess how the Tan's contact $C$ is affected by disorder at the lowest order here considered throughout the BCS-BEC crossover.
The importance of the contact, which is a measure of the number of pairs of fermions in the two spin states with small separations \cite{Tan-2008,Braaten-2012}, stems from 
the fact that it connects a number of universal relations involving several properties of a system with short-range dynamics.
For instance, the asymptotic behavior of the fermionic distribution $n(k)$ (per spin component) is characterized by the relation 
\begin{equation}
C = \lim_{k \rightarrow \infty} k^{4} \, n(k) 
\label{C-from-n_k}
\end{equation} 
\noindent
where $k = |\mathbf{k}|$.

These properties hold under quite general conditions even in the presence of an external potential, although the numerical value of the contact will depend on the specific form of external potential (for given inter-particle coupling and temperature).
In the present case of a disordered potential with the form (\ref{impurity-potential}) specified by the two parameters $\gamma$ and $p_{0}$, it may be physically more relevant to assess how the contact depends on the wave vector $p_{0}$ that characterizes the correlation length of the disorder rather than on the strength $\gamma$.
We shall then be concerned with the dependence of the contact $C$ on $p_{0}$ at finite temperature once the asymptotic regime (\ref{C-from-n_k}) has been reached for sufficiently large values of $k$, along the lines explored in Ref.\cite{PPPS-2010} for the clean case.

Alternatively, the contact $C$ can be conveniently obtained using the definition $C = (m \Delta_{\infty})^{2}$ in terms of the high-energy scale $\Delta_{\infty}$ introduced in Ref.\cite{PPS-NP-2009}, which is in turn calculated through the relation:
\begin{equation}
\Delta_{\infty}^{2} = \int \! \frac{d\mathbf{q}}{(2 \pi)^{3}} \, k_{B} T \, \sum_{\nu}  e^{i \Omega_{\nu} \eta} \, \Gamma(\mathbf{q},\Omega_{\nu}) \, .
\label{delta-infty_vs_Gamma}
\end{equation}
\noindent
In the present context, for the clean system we take in Eq.(\ref{delta-infty_vs_Gamma}) for $\Gamma$ the bare ladder propagator $\Gamma_{0}$ of Fig.\ref{fig-2}(c), while in the presence of disorder we consider the dressed ladder propagator defined by Eq.(\ref{dressed-ladder-propagator}) which includes the bosonic-like self-energy $\Sigma_{4i-4j}^{B}$.

\begin{figure}[t]
\includegraphics[angle=0,width=9.2cm]{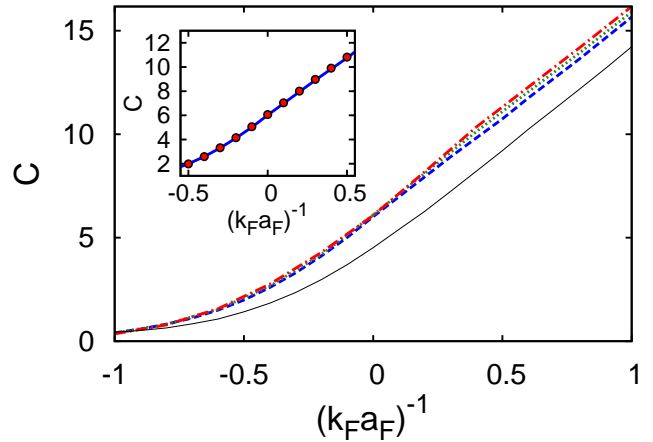}         
\caption{The contact $C$ at $T_{c}$ calculated according to Eq.(\ref{delta-infty_vs_Gamma}) vs the coupling $(k_{F} a_{F})^{-1}$, for the clean case (full line) 
              and for the disordered case with $\tilde{\gamma} =0.01$ and several values of $p_{0} / k_{F}$: 5 (dashed line), 7 (dotted line), 10 (dashed-dotted line). The inset compares 
              the values of $C$ at $T_{c}$ for the disordered case with $\tilde{\gamma} =0.01$ and $p_{0} / k_{F} = 5$, as obtained from Eq.(\ref{delta-infty_vs_Gamma}) (full line) and from the
              definition (\ref{C-from-n_k}) (dots).}
\label{fig-13}
\end{figure}

Figure \ref{fig-13} shows the coupling dependence of $C$ taken at $T_{c}$ for the clean and disordered cases, respectively, where the strength parameter $\gamma$ is held fixed while $p_{0}$ is varied. 
[Note that $C$ is dimensionless once the wave vectors are in units of $k_{F}$ ($n(k)$ is also normalized such that $ \int \! \frac{d\mathbf{k}}{(2 \pi)^{3}} \, n(k) = \frac{1}{2}$).]
We have verified numerically that in the BEC limit $(k_{F} a_{F})^{-1} \gg +1$ the relative increase $\delta C / C$ of the contact with respect to the clean case coincides with the analytic estimate $(2 k_{F} a_{F})^{2} \, \tilde{\gamma} \, p_{0}/k_{F}$, obtained by combining the approximate expression for the density given by the second line on the right-hand side of 
Eq.(\ref{density-equation-BEC-limit-1}) and the value of the bosonic chemical potential $\mu_{B}$ extracted from the modified Thouless criterion (\ref{modified-Thouless-condition}).

We have also verified that the results for $C$ obtained in this way from Eq.(\ref{delta-infty_vs_Gamma}) coincide in all cases with those obtained by looking directly at the leading asymptotic behavior (\ref{C-from-n_k}) of $n(k)$ (as it was done for the clean case in Ref.\cite{PPPS-2010}). 
To this end, we have considered consistently only the fermionc self-energy $\Sigma _{\mathrm{2a+2b}}$ (with $\Gamma$ replacing $\Gamma_{0}$ 
in the presence of disorder).
An example of the comparison between the two alternative methods (\ref{C-from-n_k}) and (\ref{delta-infty_vs_Gamma}) for calculating $C$ is shown in the inset of Fig.\ref{fig-13}.
And we have also verified in this context that the further inclusion of the self-energies $\Sigma _{\mathrm{4b}}$ and $\Sigma _{\mathrm{4c}}$ does not modify the results for $n(k)$ (and thus the contact) at the leading asymptotic order in $k$. 

The main conclusion that can be drawn from Fig.\ref{fig-13} is that \emph{the effect of disorder is to enhance the value of the contact at any coupling}, thereby somewhat favoring the occurrence of pairing correlations at least at short range.
This is because the contact $C$ is also related to the short-distance behavior of the pair correlation function between fermions with opposite spins \cite{Tan-2008,Braaten-2012}:
\begin{equation}
\lim_{\rho \rightarrow 0} g_{\uparrow \downarrow}(\rho) = \frac{C}{16 \, \pi^{2} \, \rho^{2}} \, .
\label{C-vs-pair-correlation-function}
\end{equation}
\noindent
To the extent that macroscopic coherence can be maintained in the system, this enhanced effect of pairing due to disorder may also lead to an increase of the critical temperature as already discussed in sub-section III-B for the BCS side of unitarity, as well as to an increase of the pseudo-gap energy to be discussed next.

\section{IV. Characterization of the single-particle spectral function}
\label{sec:spectral-function}

For a more complete understanding about the way disorder affects a system of fermions which still preserves an underlying remnant Fermi surface (or, else, a dilute system of weakly-interacting composite bosons when the Fermi surface has collapsed), it is instructive to analyze the single-particle spectral function corresponding to these systems in terms of the relevant self-energies.

\vspace{0.1cm}
\begin{center}
{\bf A. Fermionic spectral function}
\end{center}
\vspace{0.1cm}

We first consider the evolution of the fermionic spectral function from the BCS to the BEC limits, in terms of the ``minimal'' set of fermionc self-energies that we have identified in the paragraph following Eq.(\ref{density-equation}).
In particular, in what follows we shall limit for simplicity to consider only the effects of the fermionc self-energy $\Sigma _{\mathrm{2a+2b}}$ (with $\Gamma$ replacing $\Gamma_{0}$ in the presence of disorder) and address two relevant issues that have already emerged from panel (a) of Fig.\ref{fig-11}, one on the BCS and the other one the BEC side of unitarity:
\vspace{0.1cm}

\noindent
(i) On the BCS side of unitarity, in the presence of disorder there occurs an increase of $T_{c}$ of a few percent with respect to the clean case, indicating
that disorder favors fermion pairing in that regime.
Correspondingly, an analysis of the \emph{pseudo-gap energy} associated with this single-particle spectral function along the lines of Ref.\cite{PPPS-2012} will help clarifying this finding. 
\vspace{0.1cm}

\noindent
(ii) On the BEC side of unitarity, it should be possible to associate the coupling, at which the inclusion of additional self-energy contributions (specifically, $\Sigma _{\mathrm{4b}}$ and $\Sigma _{\mathrm{4c}}$ in panel (a) of Fig.\ref{fig-11}) becomes irrelevant for the critical temperature, to the point where the collapse of the Fermi surface eventually occurs. 
In this context, it will be relevant to extend to the presence of disorder the analysis made in Ref.\cite{Camerino-JILA-2011} for the clean system, in order to identify the value of the Luttinger wave vector $k_{L}$ where the single-particle dispersion backbends.
The presence of a \emph{finite value} for $k_{L}$ signals, in fact, the existence of an underlying remnant Fermi surface even for an interacting (and now also disordered) Fermi gas and points correspondingly to the importance of the Fermi statistics in physical quantities.

\begin{figure}[t]
\includegraphics[angle=0,width=8.7cm]{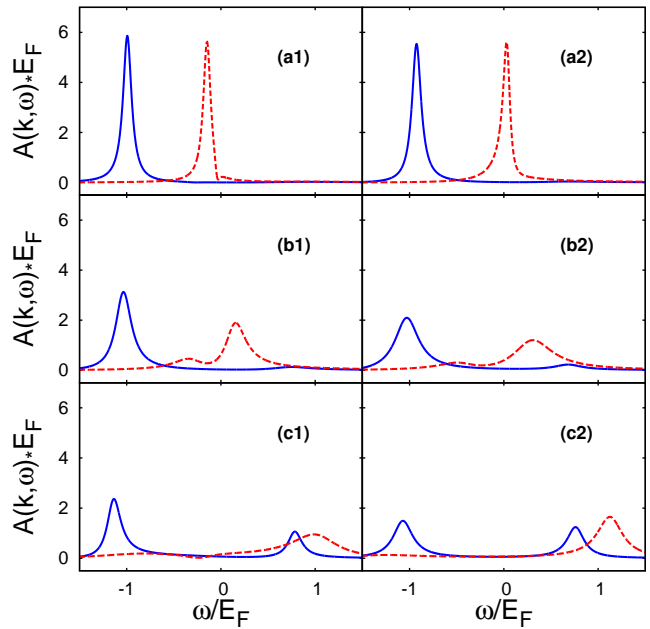}    
\caption{Single-particle spectral function $A(k,\omega)$ (in units of $E_{F}^{-1}$) vs $\omega$ (in units of $E_{F}$), for the clean (left panels) and disordered system with 
             $\tilde{\gamma} = 0.01$ (right panels), for the three couplings $(k_{F} a_{F})^{-1}=(-1.0,-0.5,0.0)$ from top to bottom. 
             The temperature is taken at the corresponding value of $T_{c}$. In each panel full lines refer to $k=0$ and broken lines to $k=k_{F}$.}
\label{fig-14}
\end{figure}

Figure \ref{fig-14} presents plots of the single-particle spectral function $A(k,\omega)$ vs the frequency $\omega$ for two characteristic values of $k$, where the analytic continuation $i \omega_{n} \rightarrow \omega + i \eta$ to real frequency $\omega$ has been taken in the expressions of the self-energy to obtain $A(k,\omega)$ \cite{footnote-Pade}.
Note how the introduction of disorder results quite generally in a shift and broadening of the peaks of $A(k,\omega)$ with respect to the clean case.
In particular, in weak coupling this shift can be attributed to the corresponding (rigid) shift of the chemical potential as reported in Fig.\ref{fig-12}.

\begin{figure}[t]
\includegraphics[angle=0,width=8.7cm]{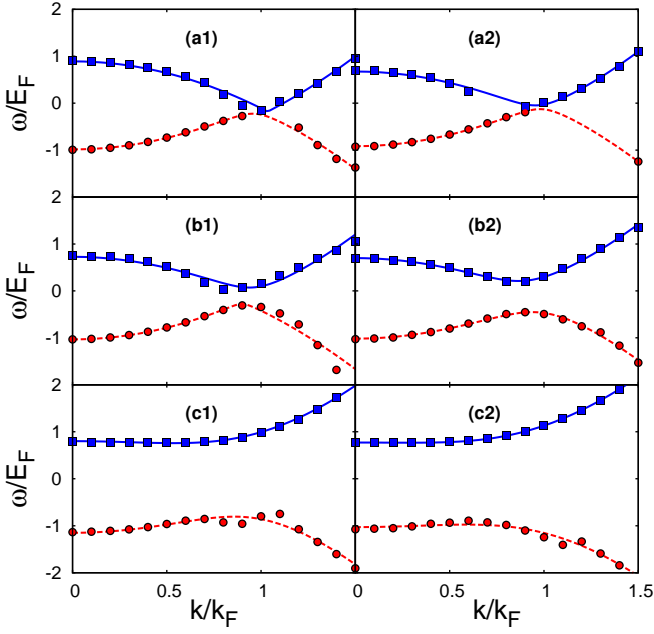}
\caption{Dispersion relations $\omega$ (in units of $E_{F}$) vs $k$ (in units of $k_{F}$), obtained by following the upper (squares) and lower (circles) peaks of the 
             single-particle spectral function for the clean (left panels) and disordered system with $\tilde{\gamma}=0.01$ (right panels) at $T_{c}$ for the same couplings considered 
             in Fig.\ref{fig-14} (from top to bottom). The lines represent the fits made following the procedure of Ref.\cite{PPPS-2012}, from which the values of the pseudo-gap energy are extracted.}
\label{fig-15}
\end{figure}

Figure \ref{fig-15} compares typical dispersion relations obtained from the single-particle spectral function $A(k,\omega)$ for the clean (left) and disordered (right) system, in the coupling interval $ -1.0 \le (k_{F} a_{F})^{-1} \le 0.0$ where an increase of $T_{c}$ is seen to occur in Fig.\ref{fig-11}(a) for the disordered with respect to the clean system.
About the middle of this coupling interval where the increase of $T_{c}$ is maximum, a wider separation between the upper and lower branches occurs in Fig.\ref{fig-15} for the
disordered with respect to the clean system.
This can be interpreted as an increase of the value of the pseudo-gap energy at $T_{c}$ and thus as a reinforcement of pairing due to disorder.

\begin{figure}[t]
\includegraphics[angle=0,width=9.0cm]{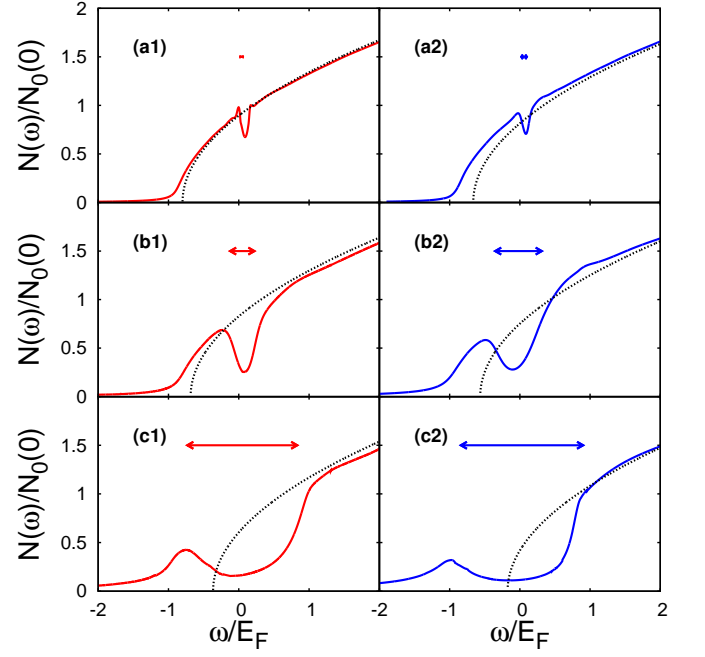}
\caption{Single-particle density of states $N(\omega)$ (in units of the non-interacting value $N_{0}(0)=m k_{F}/(2 \pi^{2})$ at the Fermi level per spin component) vs 
             $\omega$ (in units of $E_{F}$), for the clean (left panels) and disordered system with $\tilde{\gamma}=0.01$ (right panels) at $T_{c}$ for the same couplings considered 
             in Fig.\ref{fig-14} (from top to bottom). The meaning of the double arrows is explained in the text. In each panel, the dotted curve represents the 
             free-fermion-like expression $N_{0}(\omega) = \frac{m^{3/2}}{\sqrt{2} \pi^{2}} \sqrt{\omega + \mu}$ with $\omega \ge - \mu$.}
\label{fig-16}
\end{figure}

A further check on this finding, where not only the positions of the peaks of the single-particle spectral function but also their weights contribute, can be obtained by looking at the single-particle density of states:
\begin{equation}
N(\omega) =  \int_{0}^{\infty}  \! \frac{d k}{2 \pi^{2}} \, k^{2} \, A(k,\omega) \, .
\label{density_of_states}
\end{equation}
\noindent
For this quantity, the opening of a pseudo-gap at $T_{c}$ corresponds to the emergence of a depression about $\omega = 0$ with respect to the free-particle case. 
Figure \ref{fig-16} shows the plots of $N(\omega)$ obtained at $T_{c}$ for the clean (left) and disordered (right) systems, for the same three couplings on the BCS side of unitarity considered in Fig.\ref{fig-15}.
A widening of the depression about $\omega = 0$ is evident for the disordered with respect to the clean system.
In each case, the size of this depression corresponds quite well to the value of the pseudo-gap energy obtained from the plots of Fig.\ref{fig-15}, as represented by the width of the double arrow reported in each panel of Fig.\ref{fig-16}.

\begin{figure}[t]
\includegraphics[angle=0,width=6.5cm]{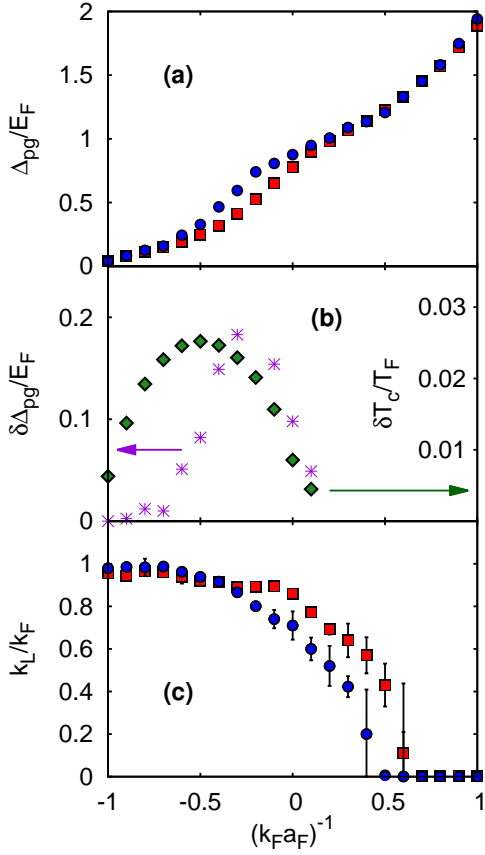}
\caption{(a) Pseudo-gap energy $\Delta_{\mathrm{pg}}$ (in units of $E_{F}$) vs the coupling $(k_{F} a_{F})^{-1}$, obtained at $T_{c}$ for the clean system (squares) 
              and the disordered system with $\tilde{\gamma}=0.01$ (circles) by an analysis of the single-particle spectral function along the lines of
              Fig.\ref{fig-15}. (b) Increase $\delta \Delta_{\mathrm{pg}}$ of the pseudo-gap energy (stars - left scale) and $\delta T_{c}$ of the critical temperature (diamonds - 
              right scale) vs the coupling $(k_{F} a_{F})^{-1}$. (c) Luttinger wave vector $k_{L}$ (in units of $k_{F}$) vs the coupling $(k_{F} a_{F})^{-1}$, obtained at the 
              corresponding values of $T_{c}$ for the clean (squares) and disordered system with $\tilde{\gamma}=0.01$ (circles).}
\label{fig-17}
\end{figure}

A comparison between the values of \emph{the pseudo-gap energy} $\Delta_{\mathrm{pg}}$ obtained in this way at $T_{c}$ for the clean and disordered system is shown in 
Fig.\ref{fig-17}(a) as a function of coupling, which has been extended to the value $(k_{F} a_{F})^{-1}=+1.0$ on the BEC side of unitarity for the purpose.
From this plot, an increase of $\Delta_{\mathrm{pg}}$ when passing from the clean to the disordered system is clearly visible approximately in the interval 
$ -0.75 \lesssim (k_{F} a_{F})^{-1} \lesssim 0$.

The increase $\delta \Delta_{\mathrm{pg}} = \Delta_{\mathrm{pg}}^{\mathrm{(dis)}} - \Delta_{\mathrm{pg}}^{\mathrm{(clean)}}$
of the pseudo-gap energy obtained from Fig.\ref{fig-17}(a) is then seen in Fig.\ref{fig-17}(b) to nicely correlate with the increase $\delta T_{c} = T_{c}^{\mathrm{(dis)}} - T_{c}^{\mathrm{(clean})}$ of the critical temperature obtained from Fig.\ref{fig-11}(a) over the relevant coupling range. 
This finding points to the conclusion that \emph{the presence of (an albeit weak) disorder appears to favor fermion pairing in this coupling regime}, thereby somewhat increasing the value of $T_{c}$ \cite{BGM-2012}.

In addition, we have extended to the BEC side of unitarity the analysis of the dispersion relations for the clean and disordered systems that was described in Fig.\ref{fig-15}, in order to identify the Luttinger wave vector $k_{L}$ where the single-particle dispersion backbends and consequently to locate the coupling value at which the collapse of the Fermi surface occurs \cite{Camerino-JILA-2011}.
The results obtained for $k_{L}$ vs $(k_{F} a_{F})^{-1}$ at $T_{c}$ are reported in Fig.\ref{fig-17}(c).
They show that the presence of disorder has the effect of moving closer to unitarity the collapse of the underlying remnant Fermi surface with respect to the clean case.
In particular, in the presence of disorder the critical coupling where $k_{L}$ vanishes is about $0.4 \div 0.5$.
This value coincides with the coupling corresponding to the empty circle in Fig.\ref{fig-11}(a) where different approximations for the fermionic self-energy all give the same result for $T_{c}$, thus signaling the emergence of a predominantly bosonic character for the system.
Past this point, Fig.\ref{fig-11}(b) shows that the critical temperature starts to depend strongly on the amount of disorder, thus implying that the system is no longer protected from the presence of disorder as soon as the underlying Fermi surface is gone.

\vspace{0.1cm}
\begin{center}
{\bf B. Bosonic spectral function}
\end{center}
\vspace{0.1cm}

It is also interesting to determine how disorder affects the spectral function of composite bosons, in the extreme BEC limit when they can be assimilated to point like-bosons for which the internal fermionic degrees of freedom become immaterial.
This case can be directly relevant to ultra-cold boson gases to the extent that their inter-particle interaction is negligible.

\begin{figure}[t]
\includegraphics[angle=0,width=6.8cm]{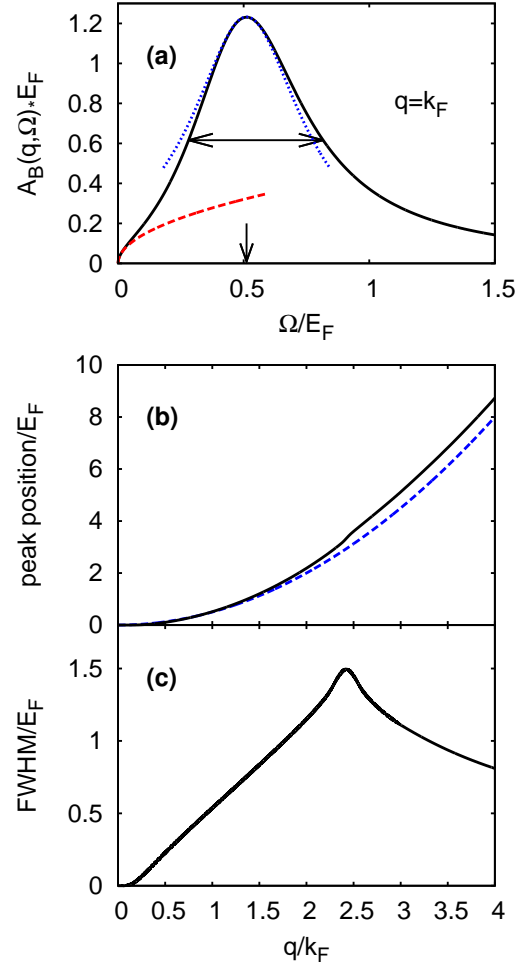}
\caption{(a) Shape of the bosonic single-particle spectral function $A_{B}(q,\Omega)$ at $T_{c}$ for a given wave vector $q$, in which the threshold behavior 
              (dashed line) and the Lorentian shape of the main peak (dotted line) are evidenced. Wave-vector dependence of (b) the position and (c) the full-width at half-maximum 
              of the main peak of $A_{B}(q,\Omega)$. The meaning of the two curves in panel (b) is explained in the text.}
\label{fig-18}
\end{figure}

Accordingly, we have calculated the bosonic self-energy $\Sigma_{B}(\mathbf{q},\Omega_{\nu})$ given by the right-hand side of 
Eq.(\ref{bosonic-self-energy-4i-j-approximate-final}) with $\tilde{\gamma}=0.01$ and $p_{0} = 5 k_{F}$, in which the bosonic chemical potential $\mu_{B}$ has been set to zero to enforce the critical condition at $T_{c}$.
(We maintain the fermionic units $k_{F}$ and $E_{F}$ also for bosonic quantities, in order to keep comparable values for the dimensionless disorder parameter $\tilde{\gamma}$.)
The analytic continuation $i \Omega_{\nu} \rightarrow \Omega + i \eta$ to real frequency $\Omega$ has then been taken to obtain the bosonic single-particle spectral function $A_{B}(q,\Omega)$, whose characteristic shape is shown in Fig.\ref{fig-18}(a) for the typical value $q = k_{F}$.
Note, in particular, the occurrence of a $\sqrt{\Omega}$ behavior at threshold (as evidenced by the dashed line) and the presence of a Lorentian form (dotted line) which well approximates the shape of the main peak locally.
The position of this peak is indicated by the vertical arrow and its full-width at half-maximum by the double horizontal arrow.

This calculation has then been repeated over an extended range of $q$ (keeping the same values $\tilde{\gamma}=0.01$ and $p_{0} = 5 k_{F}$), and has resulted in the $q$-dependence for the position and width of the peak shown respectively in Figs.\ref{fig-18}(b) and \ref{fig-18}(c).
In particular, Fig.\ref{fig-18}(b) compares the position of the peak in the presence (full line) and in the absence (dashed line) of disorder, from which one verifies that for large $q$ their difference equals $-\mu_{B}$ where $\mu_{B} = \mathrm{Re} \! \left\{ \! \Sigma_{B}(0,0) \! \right\} = - 16 \tilde{\gamma} E_{F} p_{0}/k_{F}$ is the value of the bosonic chemical potential at $T = T_{c}$ in the presence of disorder.
The width of the peak reported in Fig.\ref{fig-18}(c), on the other hand, shows a linear increase up to about the value $q = p_{0}/2$, and then decreases to zero like $q^{-1}$
for $q \gtrsim p_{0}/2$ when the bosons are eventually no longer damped by disorder.
The sizable value of this width in the presence of disorder contrasts with the delta-function shape of $A_{B}(q,\Omega)$ for non-interacting bosons in the absence of disorder, indicating again that disorder can affect bosons in a considerable way even when it is treated at the lowest order.

\section{V. Concluding remarks}
\label{sec:conclusions}

In this paper, we have presented a systematic study of the effects of disorder due to random impurities on the BCS-BEC crossover, whereby the strength of the inter-particle attraction between fermions of two different species is varied from the weak (BCS) to the strong (BEC) coupling regime.

The effects of disorder have been treated at the lowest significant order in terms of a diagrammatic approach, with the purpose of identifying the most relevant diagrams not only in the BCS and BEC limits but also in the intermediate-coupling regime about unitarity.

In this way, we have been able to follow the evolution of the effects of disorder, from a system of non-interacting fermions in the (extreme) BCS limit to a system of non-interacting composite bosons made up of fermion pairs in the (extreme) BEC limit.
In the process, the Fermi surface, that underlies the fermionic system and largely protects it from the effects of disorder, gets progressively washed out, leaving eventually the composite bosons to be much affected by the presence of disorder.
As a consequence, in future work it will be quite compelling to include the effects of disorder at higher order especially on the BEC side of unitarity.

The identification of the most relevant diagrams at the lowest order in the disorder has been done by performing accurate numerical calculations over a wide coupling window about unitarity where they are most feasible. 
This numerical analysis has also been complemented by analytic estimates of the diagrams at more extreme couplings outside this window.
The resulting crossed information between the numerical and analytical procedures has been important to arrive at a definite conclusion about the selection of a minimal set of relevant diagrams, which is sufficient to retain at the order here considered.

Numerical results for the simultaneous dependence on coupling and disorder of the critical temperature, the chemical potential, and the Tan's contact have been reported, quantities that can all be obtained at the single-particle level.
Significant features have then been extracted from these numerical results, by correlating them with a parallel analysis on the single-particle spectral function.
The increase found for the critical temperature in the presence of disorder on the BCS side of unitarity has thus been correlated with a corresponding increase of the pseudo-gap energy, which appears as a depression in the single-particle density of states.
From this analysis we have concluded that the presence of (an albeit weak) disorder somewhat favors fermion pairing in this coupling regime.
On the BEC side of unitarity, on the other hand, we have verified that the collapse of the Fermi surface makes it irrelevant to include all possible refinements of the fermionic self-energy in the presence of disorder, leaving the way to a description of the interacting fermionic system in terms of bosonic-like degrees of freedom and rendering the system much more sensitive to the presence of disorder with a marked depression of the onset of the superfluid phase.

Future work should be directed at improving on the description of disorder, possibly resting on the minimal set of diagrams here identified at the lowest significant order in the disorder.
This would imply, for instance, to replace whenever relevant the single impurity line by a ladder of impurity lines, in a similar fashion to what is done in the theory of the metal-insulator transition \cite{LR-1985,BK-1994}.
In the context of the BCS-BEC crossover, however, the impurity ladder, too, would be affected by the collapse of the Fermi surface on the BEC side of unitarity.
On the other hand, keeping the description of the effects of the attractive inter-particle interaction throughout the BCS-BEC crossover at the level of the bare t-matrix (as we have done in the present treatment) should be sufficient even when improving on the description of disorder.

\begin{figure}[t]
\includegraphics[angle=0,width=8.3cm]{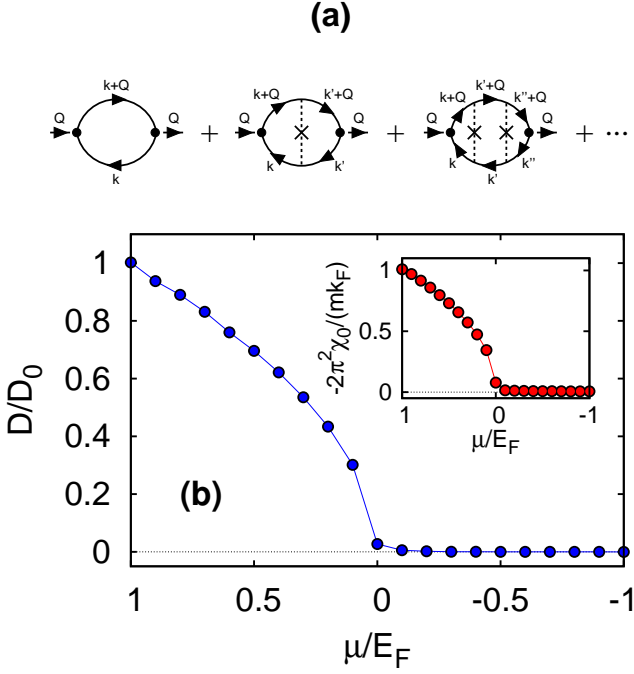}
\caption{(a) Series of diagrams for the density-density correlation function $\chi_{nn}(Q)$ that results when the fermionic self-energy is given by $\Sigma_{2a}$ of 
              Fig.\ref{fig-2}(a). Here, $Q=(\mathbf{Q},\Omega_{\nu})$ is the external four-vector. The single-particle lines are consistently calculated with the same 
              self-energy $\Sigma_{2a}$.
             (b) Function $D(\mu)$ obtained through the fit (\ref{diffusive-chi_nn}) (normalized to the non-interacting value $D_{0} = (3 \pi m \tilde{\gamma})^{-1}$).
             The inset shows the $\mu$-dependence of the pre-factor $\chi_{0}$ of the expression (\ref{diffusive-chi_nn}).}
\label{fig-19}
\end{figure}

To get a glimpse on the importance of the above arguments for future more extensive diagrammatic approaches to the BCS-BEC crossover in the presence of disorder, we may consider the calculation of the density-density correlation function $\chi_{nn}(Q)$ (per spin component) based on the diagrams of Fig.\ref{fig-19}(a), which are associated with the fermionic self-energy $\Sigma_{2a}$ in the presence of disorder.
Consistently with what we have done for the self-energy $\Sigma_{2a}$ itself (cf. Fig.\ref{fig-3}), we have obtained $\chi_{nn}(Q)$ numerically as a function of the chemical potential $\mu$.
In the present case, for a given small value of $\Omega_{\nu}$ we have fitted the numerical results for $\chi_{nn}(Q)$ vs  $\mathbf{Q}^{2}$ in terms of the diffusive form:
\begin{equation}
\chi_{nn}(Q) = \chi_{0} \, \frac{D \, \mathbf{Q}^{2}}{D \, \mathbf{Q}^{2} \, + \, \Omega_{\nu}}
\label{diffusive-chi_nn}
\end{equation}
\noindent
where $\chi_{0}$ represents the ``static'' ($\Omega_{\nu} \rightarrow 0$ and $\mathbf{Q} \rightarrow 0$) limit and $D$ plays the role of a diffusion coefficient.
The calculation has then been repeated for a few additional values of $\Omega_{\nu}$, in order to verify that the same function $D(\mu)$ is obtained in all cases.
(In this calculation, the real part of $\Sigma_{2a}$ is taken to vanish, so that the limit $p_{0} \rightarrow \infty$ can consistently be taken when calculating the diagrams of Fig.\ref{fig-19}(a).)
The function $D(\mu)$ obtained in this way is reported in Fig.\ref{fig-19}(b) and shows a characteristic $\sqrt{\mu}$ behavior for $\mu \ge 0$, vanishing accordingly when $\mu = 0$ and remaining zero for $\mu > 0$.
Akin the self-energy, additional diagrammatic contributions need then to be considered to account for the effects of disorder on the BCS-BEC crossover when dealing with the response functions from which the diffusion coefficient can be extracted.
This will be especially important when addressing the role of (weak) localization on the BCS-BEC crossover.

\vspace{0.1cm}

\begin{center}
\begin{small}
{\bf ACKNOWLEDGMENTS}
\end{small}
\end{center}
\vspace{-0.1cm}

F. P. is indebted to M. Inguscio e G. Roati for their support and interest in this work.
G. C. S. thanks R. Raimondi for discussions during the initial stage of this work.
This work was partially supported by the Italian MIUR under Contract Cofin-2009 ``Quantum gases beyond equilibrium''.
 
 \vspace{0.4cm}                                                                                                                                                                                                                                                                                                                                                                                                         
\appendix
\section{APPENDIX A: AVERAGES OVER DISORDER}
\label{sec:appendix_A}

Theoretical treatments of disorder in condensed matter usually deal with averaging over the impurity configurations by considering a Gaussian correlated (white noise) disorder with $\langle u(\mathbf{r}) u(\mathbf{r'}) \rangle = \gamma \, \delta(\mathbf{r}-\mathbf{r}')$.
Here, $u(\mathbf{r}) = \sum_{i=1}^{N} v(\mathbf{r} - \mathbf{R}_{i})$ is the potential due to $N$ impurities randomly distributed over spatial positions $\mathbf{R}_{i}$.

In the present treatment, we have adopted a truncated version of this impurity potential, taken of the form (\ref{impurity-potential}) with a finite value of the wave-vector cutoff $p_{0}$.
This is because the presence of a finite $p_{0}$ keeps also at finite values the effect of the impurities on the chemical potential, whose control is essential for driving the BCS-BEC crossover between the two BCS and BEC limits.
At the same time, this choice allows us to deal in a convenient way with the relevant diagrammatic structure in the wave-vector representation, as it was shown in Section II through a number of examples.
 
Specifically, to the form (\ref{impurity-potential}) that we have adopted there corresponds the following finite-range correlator:
\begin{equation}
\langle u(\mathbf{r}) u(\mathbf{r'}) \rangle_{p_{0}} = \frac{\gamma \, p_{0}^{2}}{2 \pi^{2} \, |\mathbf{r}-\mathbf{r}'|} \,\, j_{1}(p_{0} |\mathbf{r}-\mathbf{r}'|)
\label{finite-range-correlator-I}
\end{equation}
\noindent
where $j_{1}(z) = \frac{\sin(z)}{z^{2}} - \frac{\cos(z)}{z}$ is a spherical Bessel function.
The expression on the right-hand side of Eq.(\ref{finite-range-correlator-I}) approaches the value $\gamma \, p_{0}^{3}/(6 \pi^{2})$ in the limit
$|\mathbf{r}-\mathbf{r}'| \rightarrow 0$, presents oscillations for finite $|\mathbf{r}-\mathbf{r}'|$ which become wilder for increasing $p_{0}$, and preserves the finite area
\begin{equation}
 \int \! d(\mathbf{r}-\mathbf{r}') \langle u(\mathbf{r}) u(\mathbf{r'}) \rangle_{p_{0}} = \gamma
\label{area-finite-range-correlator-I}
\end{equation}
\noindent
irrespective of the value of $p_{0}$.

\begin{figure}[t]
\begin{center}
\includegraphics[angle=0,width=9.1cm]{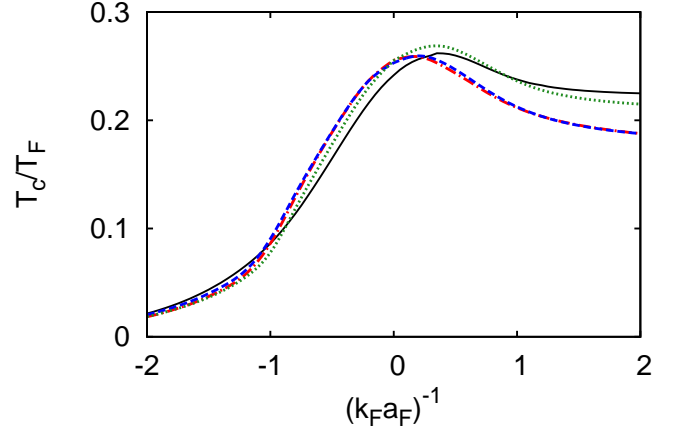}
\caption{Critical temperature (in units of the Fermi temperature) vs the coupling $(k_{F} a_{F})^{-1}$ for a fixed disorder strength $\tilde{\gamma} = 0.01$.
              Several values of the wave-vector cutoff $p_{0}/k_{F}$ are considered: $2$ (dotted line); $5$ (dashed line); $10$ (dash-dotted line). 
              The curve corresponding to a clean system is also reported for comparison (full line).}
\label{fig-20}
\end{center}
\end{figure}

A different approach was considered in Ref.\cite{Sa_de_Melo-2011} where $u(\mathbf{k})^{2}$ was taken of the smooth form 
$\gamma \exp \{- \ell_{d} \mathbf{k}^{2}/2 \}$, to which there corresponds the correlator:
\begin{equation}
\langle u(\mathbf{r}) u(\mathbf{r'}) \rangle_{\ell_{d}} = \frac{\gamma}{(2 \pi)^{3/2} \, \ell_{d}^{3}} \,\, e^{- \frac{|\mathbf{r}-\mathbf{r}'|^{2}}{2 \, \ell_{d}^{2}}} \, .
\label{finite-range-correlator-II}
\end{equation}
\noindent
The expressions (\ref{finite-range-correlator-I}) and (\ref{finite-range-correlator-II}) can be compared in the limit $|\mathbf{r}-\mathbf{r}'| \rightarrow 0$.
Assuming the same value for $\gamma$, one obtains $p_{0} \ell_{d} = \left( 9 \pi / 2 \right)^{1/6} \simeq 1.555$.
In both cases, the impurity potential is assumed not to sustain bound states.

Note that, for the details of disorder to be irrelevant, one has to require that $2 \pi / k_{F} \gg \ell_{d} \simeq 3 / (2 \, p_{0})$, thus implying that 
$p_{0}/k_{F} \gg 1$.
With our choice (\ref{impurity-potential}) of the impurity potential, we have verified numerically that when $p_{0}/k_{F}$ lies in the range between $5$ and $10$ the results for the critical temperature across the BCS-BEC crossover are rather stable for a given value of $\gamma$.
As an example, in Fig.÷\ref{fig-20} we show these results for the value $\tilde{\gamma} = 0.01$ where $\tilde{\gamma}= \gamma N_{0}/E_{F}$ is the dimensionless disorder strength.

\begin{figure}[t]
\begin{center}
\includegraphics[angle=0,width=8.8cm]{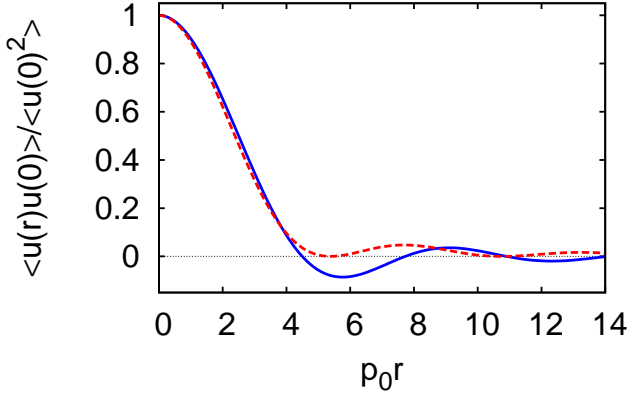}
\caption{Comparison between the radial profiles of the correlators (\ref{finite-range-correlator-I}) (full line) and (\ref{speckle-correlator}) (dashed line) when the conditions 
(\ref{conditions_vs_speckle}) hold. The normalization of the correlators is also consistent with the conditions (\ref{conditions_vs_speckle}).}
\label{fig-21}
\end{center}
\end{figure}

It also relevant to find a way to compare the correlator (\ref{finite-range-correlator-I}), which we have used in our theoretical treatment of the effects of a weak disorder on the BCS-BEC crossover, with the correlator associated with a speckle potential \cite{speckle-2007}, which has been utilized thus far in the experiments with ultra-cold (Bose as well as Fermi) atoms \cite{Aspect-2008,Inguscio-2008,De_Marco-2011,Aspect-2012}.
To this end, we compare in Fig.÷\ref{fig-21} the spatial profiles of the finite-range correlator (\ref{finite-range-correlator-I}) and of the correlator corresponding to the speckle disorder, namely, 
\begin{equation}
\langle u(\mathbf{r}) u(\mathbf{r'}) \rangle_{\mathrm{s}} = \, V^{2} \,  
\left[ \frac{\sin \left( \frac{|\mathbf{r}-\mathbf{r}'|}{\ell_{s}} \right)}{ \frac{|\mathbf{r}-\mathbf{r}'|}{\ell_{s}}} \right]^{2} \, ,
\label{speckle-correlator}
\end{equation}
\noindent
in such a way that the correlators (\ref{finite-range-correlator-I}) and (\ref{speckle-correlator}) have the same value when $\mathbf{r} = \mathbf{r}'$ and enclose the same volume up to their respective first nodes in the variable $\mathbf{r} - \mathbf{r}'$.

We obtain the condition $\gamma \, p_{0}^{3}/(6 \pi^{2}) = V^{2}$ by equating the values of the correlators at $\mathbf{r} = \mathbf{r}'$, and the condition
$1.6754 \gamma = 2 \pi^{2} V^{2} \ell_{s}^{3}$ by equating the volumes enclosed up to the respective first node (which lies at $|\mathbf{r}-\mathbf{r}'|=4.4934/p_{0}$
for the correlator (\ref{finite-range-correlator-I}) and at $|\mathbf{r}-\mathbf{r}'|=\pi \ell_{s}$ for the correlator (\ref{speckle-correlator})).
This yields:
\begin{equation}
p_{0} = \frac{1.713}{\ell_{s}} \,\,\,\,\,\, \mathrm{and} \,\,\,\,\,\, \gamma = 11.78 \, V^{2} \, \ell_{s}^{3} 
\label{conditions_vs_speckle}
\end{equation}
\noindent
which in dimensionless units correspond to
\begin{equation}
\frac{p_{0}}{k_{F}} = \frac{1.713}{k_{F} \ell_{s}} \,\,\,\,\, \mathrm{and} \,\,\,\,\, \tilde{\gamma} = 0.298 \left(\frac{V}{E_{F}}\right)^{2}\,
\left(k_{F} \, \ell_{s} \right)^{3} .
\label{conditions_vs_speckle-dimensionless}
\end{equation}

We have already specified that, in practice, $p_{0}/k_{F}$ should not be taken smaller than $5$.
The maximum value of $\tilde{\gamma}$, on the other hand, is limited by the requirement that in the BCS limit the shift of the fermionic chemical potential due to disorder (as given by the real part of the self-energy (\ref{approximate-self-energy-BCS})) should not exceed, say, $10 \%$ of the value of the Fermi energy.
This gives: 
\begin{equation}
\frac{m \, \gamma \, p_{0}}{\pi^{2} E_{F}} = 2 \, \frac{\gamma}{E_{F}/N_{0}} \, \frac{p_{0}}{k_{F}} \simeq 10 \, \tilde{\gamma} \lesssim 10^{-1}
\label{condition_on_gamma}
\end{equation}
\noindent
from which we obtain $\tilde{\gamma} \lesssim 10^{-2}$.

Experimentally, with the speckle disorder the values of the parameters $k_{F} \, \ell_{s}$ and $V/E_{F}$ range approximately in the 
intervals $0.6 \div 3.6$ and $1 \div 10$, in the order \cite{Aspect-2008,Inguscio-2008,De_Marco-2011,Aspect-2012}.
From the above arguments, this corresponds to an upper value of about $3$ for the parameter $p_{0}/k_{F}$ and at the same time to a lower value of about
$6.5 \times 10^{-2}$ for the parameter $\tilde{\gamma}$ of the broadened Gaussian potential.
These values can still be considered within the boundaries of our approach.




\begin{thebibliography}{99}
\bibitem{footnote-condition} The condition $m \gamma p_{0} \ll E_{F}$ can be interpreted as $k_{F} \ell \gg 2 \pi p_{0}/k_{F}$ where $\ell$ is the mean free path.
                                             In practice, we have taken $p_{0}/k_{F} \simeq 5$ in most numerical calculations (with the exception of those reported in
                                             Figs.÷\ref{fig-13} and \ref{fig-20}).
\bibitem{footnote-additional_diagrams} We have omitted from the present study explicit consideration to decorations by disorder that connect repeated 
                                                               self-energy diagrams of the type 2(b), since they are expected to be negligible both in the BCS and BEC regimes, 
                                                               in line with the considerations made about diagrams 4(d)-4(h).
\bibitem{footnote-xi_pair} The condition $|\mu| \gg p_{0}^{2}/(2m)$ in the BEC limit is equivalent to the statement that the size $\xi_{\mathrm{pair}}$ of the    
                                         composite bosons must be much smaller than the length $p_{0}^{-1}$ over which the disorder potential is correlated, since 
                                         $\xi_{\mathrm{pair}} \simeq a_{F}/\sqrt{2}$ and $|\mu| \simeq (2 m a_{F}^{2})^{-1}$ in this limit. Under these circumstances, the 
                                         composite bosons behave under the effect of disorder as they were point-like bosons.
\end{thebibliography}
\end{document}